\documentclass[ aps, twocolumn, nofootinbib, showkeys, notitlepage, noeprint, superscriptaddress]{revtex4-2}

\usepackage{natbib}

\usepackage{graphicx}
\usepackage{dcolumn}
\usepackage{bm}
\usepackage{amsmath}
\usepackage{float}
\usepackage{multirow}
\usepackage{slashed}
\usepackage{braket}
\usepackage{xcolor}
\usepackage{physics}
\usepackage{multirow}
\usepackage{gensymb}
\usepackage[colorlinks=true, pdfstartview=FitV, bookmarks=true, bookmarksnumbered=true, breaklinks]{hyperref}
\usepackage{mathtools,braket}

\usepackage{lipsum}  
\usepackage{color}
\definecolor{blue}{rgb}{0.0, 0.0, 1.0}
\definecolor{red}{rgb}{1.0, 0.0, 0.0}
\definecolor{royalblue}{rgb}{0.0, 0.14, 0.4}
\hypersetup{linkcolor=blue, citecolor=blue, urlcolor=blue}

\def\orcid#1{\kern .08em\href{https://orcid.org/#1}{\includegraphics[keepaspectratio,width=0.7em]{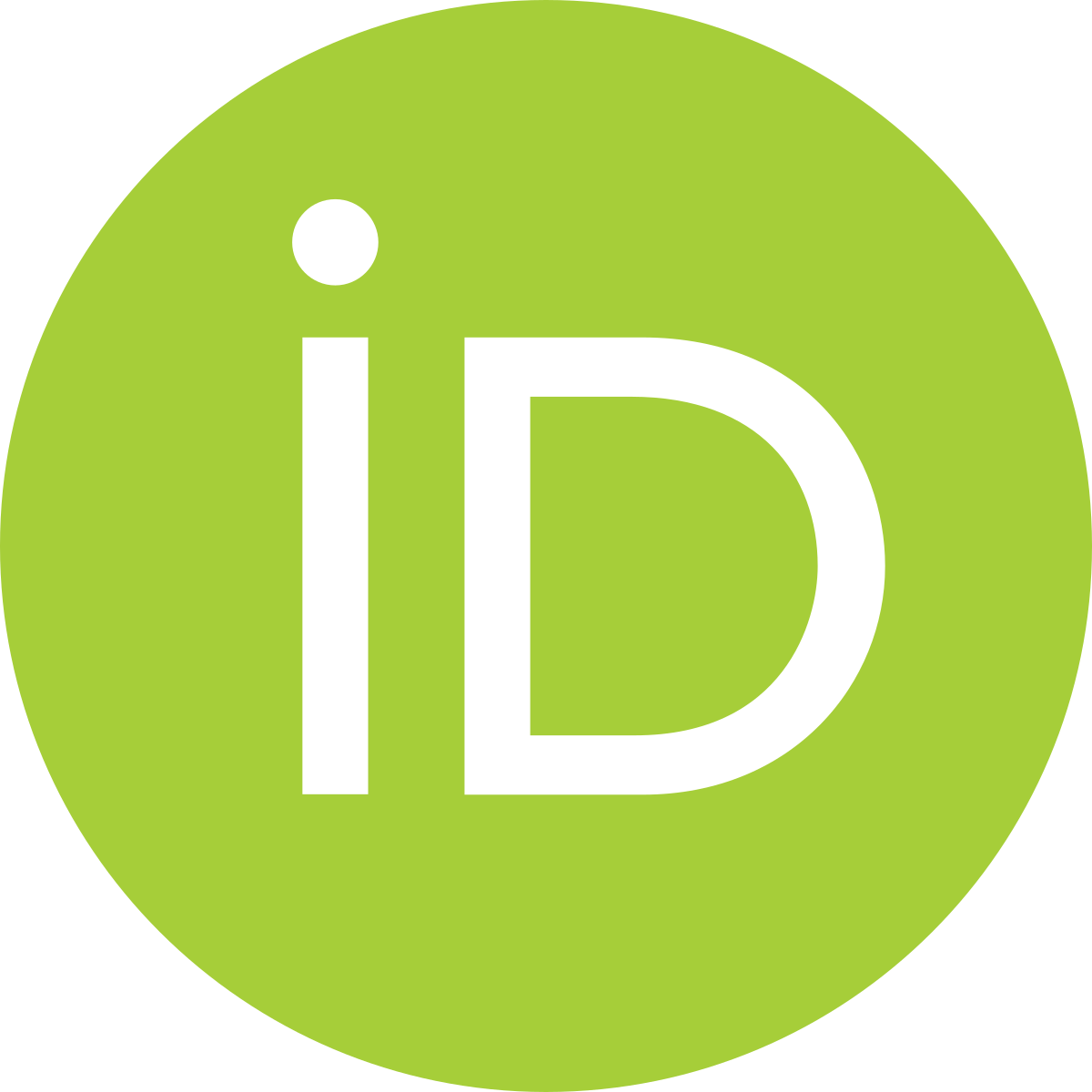}}}

\begin{document}

\title{Structure of heavy mesons in the light-front quark model}

\author{Ahmad Jafar Arifi\orcid{0000-0002-9530-8993}} 
\email{ahmad.arifi@riken.jp}
\affiliation{Few-body Systems in Physics Laboratory, RIKEN Nishina Center, Wako 351-0198, Japan}
\affiliation{Research Center for Nuclear Physics, Osaka University, Ibaraki, Osaka 567-0047, Japan}

\author{Lucas Happ\orcid{0000-0002-3893-279X}} 
\email{lucas.happ@riken.jp}
\affiliation{Few-body Systems in Physics Laboratory, RIKEN Nishina Center, Wako 351-0198, Japan}

\author{Shuhei Ohno\orcid{0009-0001-5222-9726}} 
\email{shuhei.ohno@riken.jp}
\affiliation{Few-body Systems in Physics Laboratory, RIKEN Nishina Center, Wako 351-0198, Japan}
\affiliation{Graduate School of Nanobioscience, Yokohama City University, Yokohama 236-0027, Japan}

\author{Makoto Oka\orcid{0000-0003-3206-3770}} 
\email{makoto.oka@riken.jp}
\affiliation{Few-body Systems in Physics Laboratory, RIKEN Nishina Center, Wako 351-0198, Japan}
\affiliation{Advanced Science Research Center, Japan Atomic Energy Agency, Tokai 319-1195, Japan}

\date{\today}

\begin{abstract}

We investigate the structure of ground-state heavy mesons within the light-front quark model, utilizing wave functions derived from the Single Gaussian Ansatz (SGA) and the Gaussian Expansion Method (GEM). By performing a $\chi^2$ fit to static properties such as mass spectra and decay constants, we determine the model parameters for each approach. We then compare the impacts of both methods on the light-front wave functions and structural observables.
Our analysis reveals significant differences in the distribution amplitudes (DAs) $\phi_{2;M}(x)$ near the endpoints, with GEM showing enhanced amplitudes and correct asymptotic behavior $\phi_{2;M}(x \to 1) \propto (1-x)$, consistent with perturbative QCD. This endpoint behavior is linked to the short-range (high-momentum) wave function governed by color Coulomb interaction and relativistic kinematics. GEM accurately reproduces a power-law damping $\psi_0(k \to \infty) \propto 1/k_\perp^2$, aligning with perturbative QCD predictions.
Furthermore, the electromagnetic form factors of pseudoscalar mesons in the low-$Q^2$ region fall off faster with GEM than with SGA. 
Overall, while both methods adequately describe static properties, GEM provides a more accurate description of structural properties, being more sensitive to details and asymptotic behaviors.

\end{abstract}

\maketitle

\section{Introduction} 

In Quantum Chromodynamics (QCD), the light-front dynamics~\cite{Brodsky:1997de, Dirac:1949cp, Bakker:2013cea} has emerged as a promising tool for handling relativistic effects, owing to its rational energy-momentum dispersion relation, maximal number of kinematic generators, and suppression of quantum fluctuations of the vacuum. Within this framework, the light-front quark model (LFQM)~\cite{Chung:1988mu, Ji:1992yf, Cardarelli:1994yq, Cheng:1996if}, based on light-front dynamics and the constituent quark picture, has achieved significant success in characterizing various hadron phenomenologies~\cite{Choi:1999nu, Choi:2007yu, Chen:2021ywv, Zhang:2023ypl, Schlumpf:1994bc, Hwang:2010hw, Choi:2015ywa, Ke:2013yka, Ke:2010vn, Jaus:1999zv, Jaus:1996np, Arifi:2023jfe, deMelo:2012hj}.

One of the main objectives of LFQM analyses is to derive light-front wave functions (LFWFs), from which static and structural properties of hadrons can be determined. 
Within the constituent quark model, the LFWFs can be computed by various approaches such as using a simple Ansatz for the LFWFs~\cite{Choi:2007yu,Chen:2021ywv, Zhang:2023ypl}, employing the Bethe-Salpeter amplitude~\cite{deMelo:1997cb,Raya:2021zrz,Eichmann:2021vnj,dePaula:2023ver}, diagonalizing the light-front~\cite{Li:2017mlw, Jia:2018ary} or nonrelativistic Hamiltonian~\cite{Wu:2022iiu}. 
Furthermore, there also exist other methods such as the Dyson-Schwinger method~\cite{Maris:2003vk,Eichmann:2016yit,Raya:2024ejx,Ding:2022ows}, the light-front holographic model~\cite{Brodsky:2003px}, and the light-front Nambu–Jona-Lasinio model~\cite{Naito:2004vq}.

Often, a single Gaussian Ansatz (SGA)~\cite{Choi:2007yu, Chen:2021ywv, Zhang:2023ypl}
or power-law Ansatz~\cite{Schlumpf:1994bc, Hwang:2010hw, Geng:2016pyr} is employed for the LFWFs,
whose parameters are fitted to the decay constants without considering the Hamiltonian~\cite{Cheng:1996if, Hwang:2010hw, Zhang:2023ypl}. 
Alternatively, the LFWFs can be derived from the Bethe-Salpeter amplitude on the light front~\cite{deMelo:1997cb}, where the regulator parameter is fitted to the data. 
Nevertheless, once the model parameters are well-tuned, the predictions of such models can be sufficiently consistent with the data.

In another approach, LFWFs can be obtained by directly diagonalizing the light-front Hamiltonian, as exemplified in basis light-front quantization (BLFQ)~\cite{Li:2017mlw, Jia:2018ary}. Here, basis functions are constructed as a product of longitudinal and transverse components, resulting in WFs with cylindrical symmetry~\cite{Yoshida:2016xgm}, rather than spherical symmetry. However, it should be noted that LFWFs have a nontrivial $(x,k_\perp)$ dependence and in general cannot be separated~\cite{Shi:2018zqd}. 
Due to this construction, the spherical symmetry of the WFs is not fully realized. This contrasts with LFWFs within the SGA, which inherently exhibit spherical symmetry by construction~\cite{Arifi:2022qnd, Arifi:2023uqc,Pasquini:2014ppa}.

In a different approach, the effective Hamiltonian in the instant form is constructed, and the parameters of the Ansatz WFs are determined using the variational principle. Subsequently, the WFs are mapped into LFWFs~\cite{Choi:1997iq, Choi:1999nu}. While this approach has proven successful in describing mass spectra and various observables~\cite{Choi:2015ywa, Arifi:2022pal, Dhiman:2019ddr}, the SGA has limitations in describing some data and asymptotic behavior. For instance, within the SGA, the distribution amplitudes (DAs) near the endpoints are suppressed compared to lattice QCD data~\cite{Arifi:2023jfe}, and the fall-off of the calculated transition form factor is slower compared to BaBar data~\cite{Ryu:2018egt}. It is, therefore, crucial to assess the limitations of the Ansatz by contrasting it with an approach that aims to closely resemble the eigenstate of the Hamiltonian.

One method to achieve that is the Gaussian Expansion Method (GEM)~\cite{Hiyama:2003cu, Hiyama:2012sma, Hiyama:2018ivm}, which has proven its flexibility in many systems from atomic and nuclear physics~\cite{Mitroy:2013eom}.
This method has also been applied to the non-relativistic quark model to obtain mass spectra and other observables, not only for mesons~\cite{Hao:2022vwt, Taboada-Nieto:2022igy, Martin-Gonzalez:2022qwd} but also for the baryons~\cite{Yoshida:2015tia} and multiquark systems~\cite{Hiyama:2018ukv, Meng:2020knc, Meng:2021yjr, Meng:2023for, Liu:2019vtx, Kim:2022mpa, Hu:2022zdh}.
The GEM relies on the construction of realistic WFs by utilizing Gaussian basis functions with multiple range parameters. 
This allows to approximate any shape of the WF and can be used to find the eigenstates of a given Hamiltonian.

In this article, we investigate the structure of ground-state heavy mesons within the LFQM, utilizing LFWFs obtained through both SGA and GEM. 
We focus on heavy mesons due to their suitability for probing the nonrelativistic limit of the quark model.
To accomplish this, we conduct a $\chi^2$ fitting procedure on static properties such as the mass spectra and the decay constants for each method, determining the parameters associated with the effective Hamiltonian.
We then analyze the LFWFs and other related quantities such as DAs and electromagnetic (EM) form factors to understand the distinctions between the methods and provide comparisons with the experimental and lattice QCD data.
By contrasting the results for both methods, we shed light on the structure of the heavy mesons.

Our investigation reveals that both GEM and SGA yield comparable accuracy in reproducing static properties.
However, we identify differences between the two methods for the DAs and EM form factors, which are both more sensitive to the details of the WFs. 
In particular, we observe that the DAs near the endpoints for the GEM are pronounced compared to those in SGA, exhibiting the behavior $\phi(x\to 1) \propto (1-x)$~\cite{Shi:2018zqd}. 
Also, the $S$-wave WF in the high-momentum region shows a power-law damping, $\psi(k\to \infty) \propto 1/k_\perp^2$, in line with predictions from perturbative QCD~\cite{Ji:2003yj}. 
We emphasize that the endpoint behaviors are linked to the short-range region, influenced by relativistic kinematics and Coulomb interaction.
In the low-$Q^2$ region, the EM form factors of pseudoscalar mesons fall off faster for the GEM than those for the SGA.

The article is structured as follows. In Section~\ref{sec:method}, we explain the basic components of LFQM and distinguish between SGA and GEM methods. 
Additionally, we outline the procedure for obtaining LFWFs and other properties. 
In Section~\ref{sec:result}, we discuss the numerical results obtained through both methods. 
Finally, the article concludes with Section~\ref{sec:conclusion}, summarizing our findings.

\section{Method} 
\label{sec:method}

In this section, we first describe the model Hamiltonian in the instant form. We then explain the differences between the two methods: SGA and GEM. The asymptotic behaviors of the WFs are also discussed. After that, we outline the procedure for constructing the LFWFs from the instant from WFs calculated by the two methods.
Additionally, we present the observables considered in this work within the LFQM and the fitting procedures used to determine the model parameters.

\subsection{Effective Hamiltonian}

First of all, let us consider the relativistic Schrodinger equation
\begin{eqnarray}\label{eq:schrodinger}
H_{q\bar{q}} \ket{\varPsi_{q\bar{q}}} = M_{q\bar{q}} \ket{\varPsi_{q\bar{q}}},
\end{eqnarray}
where $M_{q\bar{q}}$ and $\varPsi_{q\bar{q}}$ are the eigenvalue and eigenfunction for mesons made of quark and antiquark.
The Hamiltonian in the instant form is given by
\begin{eqnarray} \label{eq:Hqq}
H_{q\bar{q}} = H_0 + V_{q\bar{q}}
\end{eqnarray}
with the usual nonrelativistic kinetic energy replaced by the relativistic one as
\begin{eqnarray} \label{eq:KE}
 H_0 = \sqrt{m_q^2 + \bm{k}^2} + \sqrt{m_{\bar{q}}^2 + \bm{k}^2},
\end{eqnarray} 
with the quark (antiquark) mass $m_q(m_{\bar{q}})$ and the relative momentum $\bm{k}=(k_z,\bm{k}_\perp)$. 
However, in this case, we cannot factorize the c.m. motion and we can only work in the rest frame of mesons $(P_\text{c.m.}=0)$.
This approach is often referred to as the relativized quark model.

In this work, we focus on the ground state spin 0 (pseudoscalar, P) and spin 1 (vector, V) heavy mesons, containing one or two heavy quarks, $c$ or $b$.
We adopt the QCD-motivated interquark potential $V_{q\bar{q}}$, which consists of the sum of the confining, color Coulomb, and hyperfine potentials, as given by
\begin{eqnarray}\label{eq:potential}
V_{q\bar{q}} &=& a + br -\frac{4\alpha_s}{3r} 
+ \frac{32\pi \alpha_s \tilde{\delta}^3(r) }{9m_q m_{\bar{q}}} (\bm{S}_q \cdot \bm{S}_{\bar{q}}),
\end{eqnarray}
where the term $\Braket{\bm{S}_q \cdot \bm{S}_{\bar{q}} }$ yields the values of $1/4$ and $-3/4$ for the vector and pseudoscalar mesons, respectively.
Although the tensor potential may contribute via the $S$- and $D$-wave mixing, its contribution is known to be weak in the quark model~\cite{Godfrey:1985xj}, and therefore we neglect it in the present work. 

Overall the model has eight parameters, four of which are the quark masses ($m_q,m_s,m_c,m_b)$. The $a$ and $b$ are parameters for the confining potential, and $\alpha_s$ is the strong running coupling, taken as a constant parameter.  
Here, we smear the spin-spin interaction with a Gaussian function as
\begin{eqnarray}
    \tilde{\delta}^3(r) = \frac{\tilde{\Lambda}^3}{\pi^{3/2}} \mathrm{e}^{-\tilde{\Lambda}^2 r^2}.
\end{eqnarray}
where $\tilde{\Lambda}$ determines the strength of the smearing effect, and we introduce a phenomenological quark mass dependence as $\tilde{\Lambda} = \Lambda \mu_q^{1/2}$ with the reduced mass $1/\mu_q = 1/m_q + 1/m_{\bar{q}}$. 
This accommodates the fact that the size gets smaller for heavier mesons. 
Not only the hyperfine splitting but also the decay constants are affected due to such a dependence~\cite{Choi:2015ywa}. In this work, the model parameters in the Hamiltonian are determined through a $\chi^2$ fit, which will be explained in Section~\ref{sec:fit}. 

\subsection{SGA and GEM}

In this study, we consider two approaches to solve Eq.~\eqref{eq:schrodinger} with the Hamiltonian in the instant form: (i) single Gaussian Ansatz (SGA) and (ii) Gaussian expansion method (GEM), whose resulting WFs are both mapped subsequently into the LFWFs.
Here, we use the same form of the model Hamiltonian for both SGA and GEM but allow a different set of parameters for each. 
In the following, we explain the two methods in more detail.

\subsubsection{Single Gaussian Ansatz} 

The single Gaussian Ansatz (SGA) relies on making an Ansatz for the meson WFs in the form of a single Gaussian function~\cite{Choi:1997iq, Choi:1999nu}. 
The trial WF in position space is given by
\begin{eqnarray}
    \psi(\bm{r}) = \frac{(2\nu)^{3/4}}{\pi^{3/4}} \mathrm{e}^{-\nu r^2}, 
\end{eqnarray}
and the WF in the momentum space, obtained through the Fourier transformation, is given by
\begin{eqnarray}
    \psi(\bm{k}) = \frac{1}{ (2\pi\nu)^{3/4}} \mathrm{e}^{-{k}^2/ 4\nu}.
\end{eqnarray}
where the $S$-wave spherical harmonic $Y_{00} = 1/\sqrt{4\pi}$ is already included.
It is important to note that $\bm{r}$ represents the relative coordinate between the quark and anti-quark.
While the Hamiltonian parameters are adjusted by fitting the mass spectra, 
the Gaussian parameter $\nu$ for each meson is determined through the variational principle $\partial M_{q\bar{q}}/\partial\nu=0$.

Once the model parameters of the Hamiltonian are well-tuned, the predictions from this approach can reasonably agree with experimental data for a wide range of observables. However, it is important to note that a single Gaussian is an eigenfunction of the harmonic oscillator (HO) potential, not a general Hamiltonian. Because of this, the SGA does not accurately reflect the correct shape and asymptotic behavior of the WF for the Hamiltonian in Eq.~\eqref{eq:Hqq}. Therefore, the eigenstate for a given Hamiltonian is expected to deviate from a single Gaussian shape, even though the size of the WF after fitting can be similar.

\subsubsection{Gaussian Expansion Method}

To overcome the limitations of the SGA, we employ the Gaussian expansion method (GEM)~\cite{Hiyama:2003cu, Hiyama:2012sma, Hiyama:2018ivm} to solve the relativistic Schrödinger equation in Eq.~\eqref{eq:schrodinger}. This method can be understood as a generalization of the SGA, as we increase the number of Gaussian functions until the necessary accuracy for approximating the solution of Eq.~\eqref{eq:schrodinger} is achieved. Despite its similarity, it is important to note that GEM is conceptually different from the use of an Ansatz, which does not aim to find the eigenstate of the Hamiltonian.

In this method, we expand the WF in terms of a set of Gaussian basis functions, $\phi^{G}_n$, each with a different Gaussian parameter $\nu_n$ as
\begin{eqnarray} \label{eq:GEM}
     \psi &=& \sum_{n=1}^{n_\mathrm{max}} c_{n}   \phi_{n}^{\mathrm{G}}, 
\end{eqnarray}
where $c_n$ represents the expansion coefficient.
Following the notation in Ref.~\cite{Hiyama:2003cu}, the Gaussian basis function in position space is given by
\begin{eqnarray}
    \phi^{\mathrm{G}}_n(\bm{r})  =\frac{(2\nu_n)^{3/4}}{\pi^{3/4}} \mathrm{e}^{-\nu_n r^2},
\end{eqnarray}
and the basis function in the momentum space, obtained through the Fourier transformation, is given by
\begin{eqnarray}
	\phi_{n}^{\mathrm{G}} (\bm{k}) &=& \frac{1}{ (2\pi\nu_n)^{3/4}} \mathrm{e}^{-{k}^2/ 4\nu_n}.
\end{eqnarray}

In the GEM, it is necessary to determine two sets of parameters: the coefficients $c_n$ and the Gaussian parameters $\nu_n$.
To obtain the $c_n$ expansion coefficients satisfying ${\partial M_{q\bar{q}}}/{\partial c_n}=0$,  we solve the generalized eigenvalue problem
\begin{eqnarray}
    \bm{H}_{q\bar{q}} \bm{c} = M_{q\bar{q}} \bm{S} \bm{c},
\end{eqnarray}
where the elements of the Hamiltonian matrix are $H_{q\bar{q},nm} = \bra{\phi_n^\mathrm{G}}\hat{H}\ket{\phi_m^\mathrm{G}}$, and the elements of the overlap matrix are $S_{nm} = \Braket{\phi_n^\mathrm{G}|\phi_m^\mathrm{G}}$.
To obtain the Gaussian parameters $\nu_n$, we consider a geometric progression~\cite{Hiyama:2003cu} 
\begin{eqnarray}
    \nu_n = \frac{1}{r_n^2},\qquad r_n = r_1 a^{n-1}, \qquad a = \left(\frac{r_\mathrm{max}}{r_1}\right)^{\frac{1}{n_\mathrm{max}-1}}
\end{eqnarray}
which reduces them to only two ($\nu_1$, $\nu_{n_\mathrm{max}}$) while keeping a high accuracy of the calculation.
Those two parameters are optimized using the Optim.jl package~\cite{Mogensen2018} to satisfy
\begin{eqnarray}
    \frac{\partial M_{q\bar{q}}}{\partial \nu_1} = \frac{\partial M_{q\bar{q}}}{\partial \nu_\mathrm{max}}=0.
\end{eqnarray}
We note that the basis functions are non-orthogonal $S_{nm}\neq \delta_{nm}$ and the states are normalized as
\begin{eqnarray}
  \braket{\psi}{\psi} = \sum_{n,m} c_n^* S_{nm}c_m = 1.
\end{eqnarray}

\subsection{LFWFs and Observables}

\subsubsection{Mass Spectra}

In both SGA and GEM, we determine the mass spectra of the ground state heavy mesons by solving Eq.~\eqref{eq:schrodinger} in their rest frame. In this study, we calculate the mass spectra using instant form dynamics, which in line with the approach in the NRQM but incorporates relativistic kinetic energy~\cite{Hao:2022vwt, Taboada-Nieto:2022igy, Martin-Gonzalez:2022qwd}. 
This method contrasts with BLFQ~\cite{Li:2017mlw}, where mass spectra are directly derived from LFWFs.

\subsubsection{LFWFs}

After diagonalizing the Hamiltonian, we need to map the obtained WFs into the LFWFs~\cite{Choi:2015ywa, Wu:2022iiu}.
For that we proceed in the following way
\begin{enumerate}
    \item The position-space WF $\psi(\bm{r})$ is first transformed into the momentum-space WF $\psi(\bm{k})$.
    \item The radial part of $\psi(\bm{k})$ can be supplied to LFWFs $\Phi(x,\bm{k}_\perp)$ via the $k_z\to x$ variable transformation. Note that the Jacobian factor ${\partial k_z}/{\partial x}$ is necessary to maintain the rotational symmetry.
    \item The spin and orbital part of the LFWFs are obtained via the interaction-independent Melosh transformation~\cite{Melosh:1974cu}.
\end{enumerate}
Before proceeding further, we make some remarks about this mapping procedure.
In our relativized quark model, we replace the kinetic energy term by the relativistic one. 
For fully relativistic formulation based on the Dirac equation, the effects of the small components induce (spin-dependent) relativistic corrections that are important in the mass spectrum. 
In the non-relativistic formulation, instead, we include the spin-spin, spin-orbit and tensor interactions explicitly in the Hamiltonian. 
Although the relativistic effects are not fully included in the WF, our spectrum reproduces the observed one by adjusting the Hamiltonian parameters as we will see later. 
Then we map the WF to LFWF by the Melosh transformation~\cite{Melosh:1974cu}, which is independent of the interaction and consistent with the Bakamjian-Thomas construction~\cite{Bakamjian:1953kh}. 
It is not easy to quantify the ambiguity coming from this approximation in the final results, but our approach can be justified by comparing the results with data.
Thus, this approach provides a well-controlled connection between the WF and LFWF. 

In the following, we provide a more detailed demonstration of the mapping.
The LFWFs are expressed in terms of the Lorentz invariant internal variables 
\begin{eqnarray}
   x_i &=& p^+_i /P^+, \\
   \bm{k}_{\perp i} &=& \bm{p}_{\perp i} - x_i \bm{P}_{\perp},
\end{eqnarray}
where $P^\mu = (P^+,P^-,\bm{P}_\perp)$ and $p^\mu_i$ denote the four-momentum of the meson and the $i$-th constituent quark, respectively.
Here we define the longitudinal momentum fraction $x \equiv x_q$ and the transverse momentum ${\bf k}_\perp \equiv {\bf k}_{\perp q}$.

We perform the variable transformation $(k_z, \bm{k}_\perp) \to (x,\bm{k}_\perp)$. 
In this case, $x$ can be related with $k_z$ as
\begin{eqnarray}\label{kz}
    x = \frac{ E_q - k_z}{E_q + E_{\bar{q}}}, \qquad  1 - x = \frac{E_{\bar{q}} + k_z}{E_q + E_{\bar{q}}},
\end{eqnarray}
where $E_i = \sqrt{m^2_i + \bm{k}^2}$.
Therefore, $k_z$ can be written as
\begin{eqnarray}\label{eq:k_z}
    k_z = \left( x - \frac{1}{2} \right) M_0 + \frac{(m^2_{\bar{q}} -m^2_q)}{2M_0}
\end{eqnarray} 
with the so-called invariant meson mass $M_0 = E_q + E_{\bar{q}}$ expressed as
\begin{eqnarray}
	M_0^2 = \frac{\bm{k}_{\bot}^2 + m_q^2}{x}  + \frac{\bm{k}_{\bot}^2 + m_{\bar{q}}^2}{1-x}.
\end{eqnarray} 
The radial part of LFWFs is given by
\begin{eqnarray}
    \Phi(x,\bm{k}_\bot) = \sqrt{2(2\pi)^3} \sqrt{\frac{\partial k_z}{\partial x}} \psi(\bm{k}),
\end{eqnarray}
where for $\psi(\bm{k})$ we take the WF from the SGA and GEM.
The Jacobian factor is expressed as
\begin{equation}
\frac{\partial k_z}{\partial x} = \frac{M_0}{4x(1-x)} \left[ 1 - \frac{ (m_q^2 - m_{\bar{q}}^2)^2}{M_0^4} 
\right],
\end{equation}
which takes into account the variable transformation.

The spin and orbital angular momentum part, $\mathcal{R}^{JJ_z}_{\lambda_q\lambda_{\bar{q}}}$, of LFWFs is obtained via the interaction-independent Melosh transformation~\cite{Melosh:1974cu} from the spin and orbital angular momentum part of the relativistic WF in the instant form assigned to the quantum number $J^{PC}$.
The $\mathcal{R}^{JJ_z}_{\lambda_q\lambda_{\bar{q}}}$ has covariant forms as
\begin{eqnarray}
	\mathcal{R}^{JJ_z}_{\lambda_q\lambda_{\bar{q}}} &=&  \frac{1}{\sqrt{2} \tilde{M}_0} 
	\bar{u}_{\lambda_q}^{}(p_q) \Gamma_{\rm M} v_{\lambda_{\bar{q}}}(p_{\bar{q}}),
\end{eqnarray}
with $\tilde{M}_0 \equiv \sqrt{M_0^2 - (m_q -m_{\bar{q}})^2}$ and the Dirac spinors of quark $u(p_q)$ and antiquark $v(p_{\bar{q}})$. 
The vertex $\Gamma_{\mathrm{M}}$ for the pseudoscalar and vector meson (with M = V or P) is given by 
\begin{eqnarray}
\Gamma_{\rm P} &=& \gamma_5, \\
\Gamma_{\rm V} &=& -\slashed{\epsilon}(J_z) + \frac{\epsilon \cdot (p_q-p_{\bar{q}})}{ M_0 + m_q + m_{\bar{q}}},
\end{eqnarray}
where the polarization vectors $\epsilon^\mu(J_z)=(\epsilon^+, \epsilon^-,\bm{\epsilon}_{\perp})$ are defined by
\begin{eqnarray}\label{eq:7}
\epsilon^\mu(\pm 1) &=& \left( 0, \frac{2\bm{\epsilon}_\perp(\pm) \cdot \bm{P}_\perp}{P^+}, \bm{\epsilon}_\perp(\pm)\right), \nonumber\\
\epsilon^\mu(0) &=& \left(\frac{P^+}{M_0}, \frac{-M^2_0 + \bm{P}^2_\perp}{M_0P^+}, \frac{\bm{P}_\perp}{M_0}\right), 
\end{eqnarray} 
with $\bm{\epsilon}_\perp(\pm 1) = \left( 1, \pm i \right)/\sqrt{2}$. 
We note that the vertex $\Gamma_\text{P}$ can contain not only the pseudoscalar coupling, but also the pseudovector coupling. Although this pseudovector is usually not important in the low-energy regime, it can affect the asymptotic behavior of form factors as $Q^2\to \infty$~\cite{Maris:1998hc}. 

The explicit forms of the spin and orbital WFs for the pseudoscalar and vector mesons are given by
\begin{equation}
    \mathcal{R}^{00}_{\lambda_q\lambda_{\bar{q}}} =
	\frac{1}{\sqrt{2}\sqrt{\mathcal{A}^2 + \bm{k}_\perp^2}}
	\begin{pmatrix}
	k^L & \mathcal{A} \\
	-\mathcal{A}   & -k^R\\
	\end{pmatrix},\\
\end{equation}
and
\begin{equation}
\begin{aligned}
	\mathcal{R}^{11}_{\lambda_q\lambda_{\bar{q}}} &=\frac{1}{\sqrt{\mathcal{A}^2 + \bm{k}_\perp^2}}
    \begin{pmatrix}
	\mathcal{A} + \frac{\bm{k}_\bot^2}{{\mathcal D}_0}		& k^R \frac{\mathcal{M}_1}{{\mathcal D}_0}\\
	- k^R \frac{\mathcal{M}_2}{{\mathcal D}_0} & -\frac{(k^R)^2}{{\mathcal D}_0} 	\\
    \end{pmatrix},\\
	\mathcal{R}^{10}_{\lambda_q\lambda_{\bar{q}}} &=
	\frac{1}{\sqrt{2}\sqrt{\mathcal{A}^2 + \bm{k}_\perp^2}}
	\begin{pmatrix}
	k^L \frac{\mathcal{M}}{{\mathcal D}_0}	& \mathcal{A} + \frac{2\bm{k}_\bot^2}{{\mathcal D}_0}  \\
	\mathcal{A} + \frac{2\bm{k}_\bot^2}{{\mathcal D}_0}       & -k^R \frac{\mathcal{M}}{{\mathcal D}_0}	\\
	\end{pmatrix},\\
	\mathcal{R}^{1-1}_{\lambda_q\lambda_{\bar{q}}} &= \frac{1}{\sqrt{\mathcal{A}^2 + \bm{k}_\perp^2}}
    \begin{pmatrix}
	-\frac{(k^L)^2}{{\mathcal D}_0}		& k^L \frac{\mathcal{M}_2}{{\mathcal D}_0}\\
	- k^L \frac{\mathcal{M}_1}{{\mathcal D}_0} &  \mathcal{A} + \frac{\bm{k}_\bot^2}{{\mathcal D}_0}	\\
    \end{pmatrix},
\end{aligned}
\end{equation}
respectively, where $k^{R(L)}=k_x \pm ik_y$, $\mathcal{A}= (1-x)m_1 + x m_2$, 
$D_{0}= M_0 + m_q + m_{\bar{q}}$, $\mathcal{M}_1 = xM_0 + m_1$, $\mathcal{M}_2 = (1-x) M_0 + m_2$, and $\mathcal{M}=\mathcal{M}_2 - \mathcal{M}_1$.
Note that $\mathcal{R}^{JJ_z}_{\lambda_q\lambda_{\bar{q}}}$ is normalized as 
\begin{eqnarray}
\sum_{\lambda_q,\lambda_{\bar{q}}}\mathcal{R}^{JJ_z\dagger}_{\lambda_q\lambda_{\bar{q}}}\mathcal{R}^{J^\prime J_z^\prime}_{\lambda_q\lambda_{\bar{q}}} = \delta_{JJ^\prime}\delta_{J_z J_z^\prime}.
\end{eqnarray}

The LFWF of the ground state heavy meson in momentum space is therefore given by
\begin{eqnarray}~\label{eq:LFWF}
		\varPsi^{JJ_z}_{\lambda_q\lambda_{\bar{q}}}(x, \bm{k}_{\bot}) = \Phi(x, \bm{k}_\bot)
		\  \mathcal{R}^{JJ_z}_{\lambda_q\lambda_{\bar{q}}}(x, \bm{k}_\bot),
\end{eqnarray}
and normalized as
\begin{eqnarray}
 \sum_{\lambda_q,\lambda_{\bar{q}}} \int \frac{\dd x\ \dd[2] \bm{k}_\bot}{2(2\pi)^3}  \abs{ \varPsi^{JJ_z}_{\lambda_q\lambda_{\bar{q}}} (x, \bm{k}_\bot) }^2 =1.
\end{eqnarray}
Furthermore, the LFWFs are often constructed as products of $\varPsi_L(x)$ and $\varPsi_T(\bm{k}_\perp)$. 
However, this construction often lead to a breaking in spherical symmetry of the WF~\cite{Li:2017mlw}.
In our case, the LFWFs are transformed from Gaussian basis functions, thereby maintaining the spherical symmetry of the LFWFs. 
One evidence is that $\expval{\bm{k}_\perp^2=k_x^2+k_y^2}$ and $\expval{2k_z^2}$ would yield the same results when using the LFWFs in Eq.~\eqref{eq:LFWF}.
This consistency can be broken if LFWFs are constructed as $\varPsi=\varPsi_L(x)\varPsi_T(\bm{k}_\perp)$. 
Interested readers may refer to the previous study~\cite{Arifi:2022qnd} for more details.

\subsubsection{Decay Constants} 

Now we provide formulae for the decay constants of the pseudoscalar meson $f_{\rm P}$ and vector meson $f_{\rm V}^{\parallel(\perp)}$ with longitudinal and transverse polarizations.
They are defined by
\begin{eqnarray}\label{eq:17}
\bra{0} \bar{q} \gamma^\mu \gamma_5 q \ket{{\rm P}(P)} &=& i f_{\rm P} P^\mu, \\
\bra{0} \bar{q} \gamma^\mu q \ket{{\rm V}(P,J_z)} &=& f_{\rm V}^\parallel M_{\rm V} \epsilon^\mu(J_z),\\
 \bra{0} \bar{q} \sigma^{\mu\nu} q \ket{{\rm V}(P,J_z)} &=& if_{\rm V}^\perp [\epsilon^\mu(J_z)P^\nu - \epsilon^\nu(J_z)P^\mu], \quad \quad 
\end{eqnarray}
where $\epsilon^\mu (J_z)$ and $M_{\rm V}$ represent the polarization vector and the mass of the vector meson, respectively. 

The explicit form of the decay constants computed in the LFQM using the plus current $(\mu=+)$ is given by~\cite{Choi:2015ywa}
\begin{eqnarray}\label{eq:cons}
	f_{\rm M} &=& \sqrt{6}\int_0^1 \dd x\int \frac{ \dd[2] \bm{k}_\bot}{(2\pi)^3}  
	\frac{ {\Phi}(x, \bm{k}_\bot) }{\sqrt{\mathcal{A}^2 + \bm{k}_\bot^2}} ~\mathcal{O}_{\rm M}, 
\end{eqnarray}
where $\Phi(x,\bm{k}_\perp)$ is radial part of LFWFs and the operators $\mathcal{O}_{\rm M}$ read
\begin{eqnarray}
    \mathcal{O}_{\rm P} &=& \mathcal{A}, \label{eq:operator1}\\ 
    \mathcal{O}_{\rm V}^{\parallel} &=& \mathcal{A}+ \frac{2 \bm{k}_\bot^2}{D_{0}}, \label{eq:operator2}\\
    \mathcal{O}_{\rm V}^{\perp} &=& \mathcal{A}+ \frac{\bm{k}_\bot^2}{D_{0}}. \label{eq:operator3}
\end{eqnarray}
The decay constants can also be calculated using the minus $(\mu=-)$ and the transverse $(\mu=\perp)$ components of currents. 
The equivalence of the decay constants with various current components and polarizations has been demonstrated within this LFQM in a previous work~\cite{Arifi:2022qnd}. 

\subsubsection{Twist-2 Distribution Amplitudes}

Next, we focus on the twist-2 DAs which give dominant contributions in the hard exclusive processes~\cite{Lepage:1980fj}.
The DAs are derived from matrix elements that connect free space and meson states with light-like separation, i.e., $z^2=0$. 
To establish a link between the DAs and the LFWFs, we apply the condition of the equal light-front time to the light-like vector $z^\mu$ with $z^+=z_\perp=0$.
The twist-2 DAs for pseudoscalar mesons $\phi_{2, \rm P}$ computed by choosing the plus current are given by~\cite{Ball:2006wn}
\begin{eqnarray}
&\bra{0} \bar{q}(z) \gamma^+\gamma_5  q(-z) \ket{{\rm P}(P)} = i f_{\rm P} P^+ \nonumber\\
&\times {\displaystyle \int^1_0} \dd x\ {\rm e}^{i\xi P \cdot z}  \phi_{2, \rm P}(x),
\end{eqnarray}
where $\xi=2x-1$.
The twist-2 DAs for the vector mesons with longitudinal $\phi^\parallel_{2,\rm 
V}$ and transverse polarizations $\phi^\perp_{2,\rm V}$ are computed as~\cite{Ball:1998ff}
\begin{eqnarray}
&\bra{0}\bar{q}(z)\gamma^+ q(-z)\ket{{\rm V}(P,0)} = f_{\rm V}^\parallel M_{\rm V} \epsilon^+(0) \nonumber \\
& \times {\displaystyle \int^1_0} \dd x\  {\rm e}^{i\xi P\cdot z} \phi^\parallel_{2,\rm 
V}(x), \\
& \bra{0}\bar{q}(z)\sigma^{\perp+} q(-z)\ket{{\rm V}(P,\pm1)} = if_{\rm V}^\perp [\boldsymbol{\epsilon}_\perp(\pm1) P^+ \nonumber \\
&  -\epsilon^+(\pm1)\bm{P}_\perp]  {\displaystyle \int^1_0} \dd x\  {\rm e}^{i\xi P\cdot z}  \phi^\perp_{2,\rm V}(x),   
\end{eqnarray}
respectively.

\begin{figure*}[t]
	\centering
	\includegraphics[width=2.0\columnwidth]{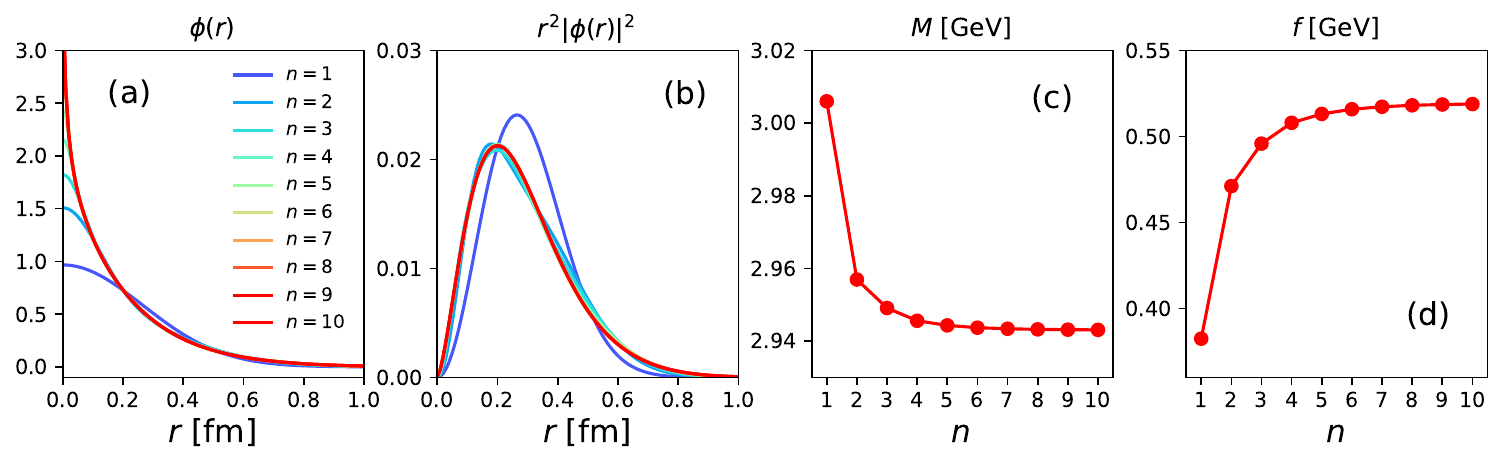}
 	\caption{\label{fig:wf} (a) WF $\phi_{\eta_c}(r)$, (b) density of WF $r^2|\phi_{\eta_c}(r)|^2$, (c) mass, and (d) decay constant of $\eta_c(1S)$ where the parameters are fixed from using the SGA ($n=1$). 
    The figures also show the results when we expand the basis function up to $n=10$ while keeping the parameters the same. Evidently, using the same parameters for $n=10$ would result in a poor prediction, especially in the decay constants, which deviate significantly up to $30\%$. Thus, it is necessary to fit the parameters independently when using the GEM ($n=10$).  } 
\end{figure*}

In the LFQM, the $\phi_{\rm M}(x)$ can be obtained by the transverse 
momentum integration of the LFWF as~\cite{Choi:2007yu}
\begin{eqnarray}\label{eq:DAs}
\phi_{2,\rm M}(x) = \frac{\sqrt{6}}{ f_{\rm M}} \int \frac{ \dd[2] 
\bm{k}_\bot}{(2\pi)^3}  \frac{ {\Phi}(x, \bm{k}_\bot) }{\sqrt{\mathcal{A}^2 + \bm{k}_\bot^2}} ~\mathcal{O}_{\rm M},
\end{eqnarray}
which are normalized as $\int_{0}^{1} {\phi}_{2,\rm M}(x)\ \dd x = 1$. 
In the model calculation, it is difficult to precisely determine the scale $\mu$, 
but it is associated with the modeling interaction.
The scale dependence of DAs can be obtained by using the QCD evolution equation~\cite{Lepage:1980fj}.

Moreover, we provide the six-lowest $\xi$-moments of the DAs, defined as
\begin{equation} \label{eq:xi}
\expval{\xi^n} = \int_{0}^{1} \dd x\ \xi^n\ \phi_{2,\rm M}(x),
\end{equation}
which can be compared with other models quantitatively.
These $\xi$ moments can be related to the Gegenbauer moments $a_n(\mu)$~\cite{Choi:2007yu}.
Although features of DAs are reflected in the first few $\xi$ moments, 
accurate parameterization to DAs of heavy-light mesons may require more moments due to a pronounced asymmetry~\cite{Serna:2020txe}. 

\subsubsection{Electromagnetic Form Factors}

Furthermore, we want to test the two methods on the transfer momentum $(Q^2)$-dependent quantities. 
For that reason, we include the EM form factor of pseudoscalar mesons
\begin{eqnarray}
    \bra{\mathrm{P}(P^\prime)}J_{\mathrm{em}}^\mu(0)\ket{\mathrm{P}(P)} =(P+P^\prime)^\mu F(Q^2),
\end{eqnarray}
where $Q^2=-q^2=-(P^\prime-P)^2$.
This EM form factor is sensitive to the quark masses and provides further tests on the details of the LFWFs. 
A more comprehensive study involving elastic and transition form factors is left for future work.

In LFQM, the EM form factor of pseudoscalar mesons is computed by using the Drell-Yan-West frame $(q^+=0)$ with $Q^2=\bm{q}_\perp^2$ and obtained by using the plus component of current $J^+$. 
The explicit expression is given by~\cite{Choi:1997iq}
\begin{equation}
F(Q^2) = e_q I^+(Q^2, m_q, m_{\bar{q}}) + e_{\bar{q}} I^+(Q^2, m_{\bar{q}},m_q),
\end{equation}
where the $e_q (e_{\bar{q}})$ is the electric charge of the quark (antiquark).
The contribution of the quark and antiquark is calculated by
\begin{eqnarray}\label{eq:emff}
	I^+(Q^2, m_q, m_{\bar{q}}) &=& \int\frac{\dd x\ \dd[2]\bm{k}_\perp}{2(2\pi)^3}
	\, \Phi(x,\bm{k}_\perp) \, \Phi^*(x,\bm{k}^\prime_\perp)  
\nonumber\\	
& &  \times 
        \frac{\mathcal{A}^2 + \bm{k}_\perp \cdot \bm{k}^\prime_\perp}{\sqrt{\mathcal{A}^2 + \bm{k}_\perp^2} 
	\sqrt{\mathcal{A}^2 + \bm{k}_\perp^{\prime 2}}}.
\end{eqnarray}
where $\Phi^*(x,\bm{k}^\prime_\perp) $ is the radial part of the final LFWF of mesons with $\bm{k}^\prime_\perp = \bm{k}_\perp + (1-x) \bm{q}_\perp$.
The form factor computed from the transverse current would also give the same results in this LFQM~\cite{Choi:2014ifm}.
Note that the EM form factor is normalized as $F(Q^2=0) = e_q + e_{\bar{q}}$, and the corresponding charge radius is computed as
\begin{eqnarray}
	\Braket{ r^2_{\rm em} } = -6 \frac{\dd F(Q^2)}{ \dd Q^2}\biggl|_{Q^2=0}.
\end{eqnarray}
In the case of heavy quarkonia such as $\eta_c$ and $\eta_b$, we only consider the contribution from one of the quarks, as otherwise the EM form factor vanishes because the contributions from the quark and antiquark cancel each other out. 
However, the EM form factors for neutral mesons such as $D^0, B^0,$ and $B_s^0$ do not vanish due to  different quark flavors.

\begin{figure*}[t]
	\centering
	\includegraphics[width=2\columnwidth]{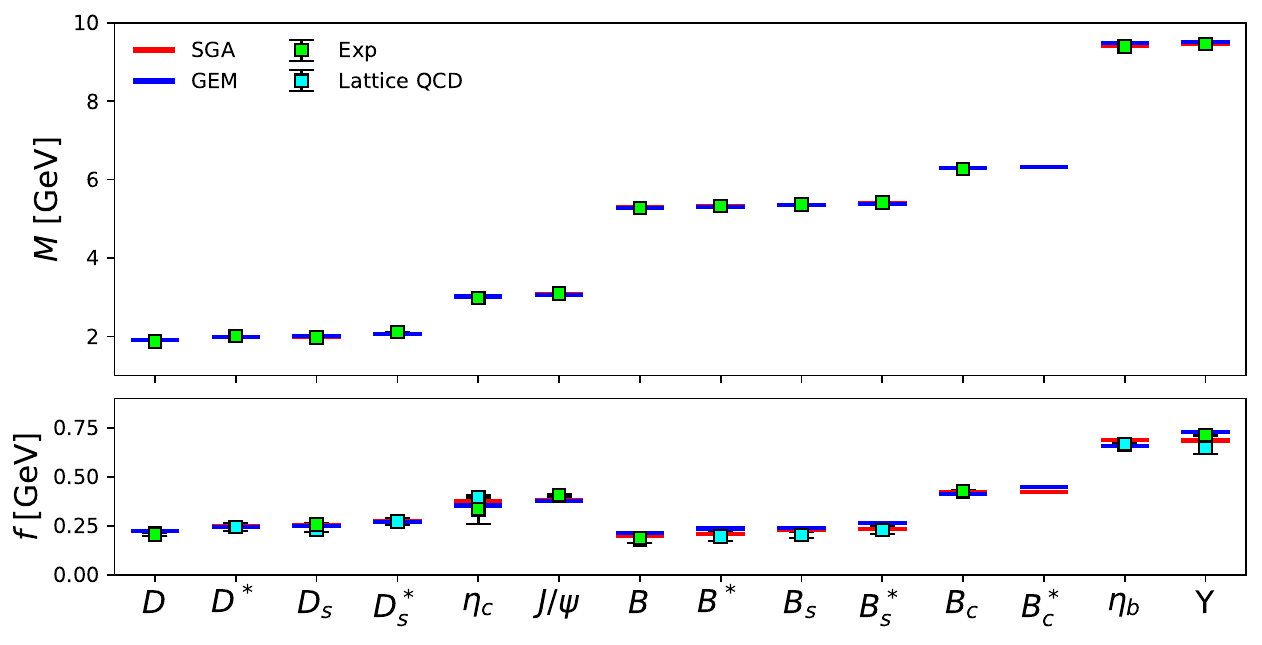}
 	\caption{\label{fig:mass} 
  Fitted mass spectra (upper panel) and decay constants (lower panel) of ground-state heavy mesons using SGA (red) and GEM (blue). In addition, we display data from experiments (light green)~\cite{ParticleDataGroup:2022pth} and lattice QCD calculations (light blue)~\cite{McNeile:2012qf, Davies:2010ip, Donald:2012ga, Colquhoun:2014ica,Hwang:2010hw}. We incorporated both observables in our fitting process.}
\end{figure*}

\begin{table*}[t]
	\begin{ruledtabular}
		\caption{ Model parameters for SGA and GEM which are independently fitted to the data by minimizing $\chi^2$. The parameters include the constituent quark masses $(m_q,m_s,m_c,m_b)$ in unit of GeV, confinement potential parameters $a$~[GeV] and $b$~[GeV$^2$], the strong coupling $\alpha_s$ which is dimensionless and the smearing parameter $\Lambda$~[GeV$^{1/2}$] with a quark mass dependence $\tilde{\Lambda}=\Lambda\mu_q^{1/2}$. Here we have $N_\mathrm{data}=32$ data points and $N_\mathrm{par}=8$ free parameters resulting in $N_\mathrm{dof} = N_\mathrm{data} - N_\mathrm{par} = 24$. We obtain $\chi^2/N_\mathrm{dof}$ is 0.95 and 0.46 for GEM and SGA, respectively. } 
		\label{tab:parameter}
		\begin{tabular}{lcccccccc}
		 Model &  $m_q$ & $m_s$ & $m_c$ & $m_b$ & $\alpha_s$ & $\Lambda$ & $a$ & $b$\\ \hline 
         SGA    & $0.2200$ & $0.3678$ & $1.6324$ & $5.0557$ & $0.4410$ & $0.9639$ & $-0.4235$ & $0.1655$ \\
         GEM    & $0.2200$ & $0.3463$ & $1.5147$ & $4.8800$ & $0.2850$ & $1.4376$ & $-0.1895$ & $0.0924$ \\
          \end{tabular}
	\end{ruledtabular}
\end{table*}

\subsection{Fitting Procedures}
\label{sec:fit}

To determine the model parameters of the Hamiltonian in Eq.~\eqref{eq:schrodinger}, we perform a $\chi^2$ fit, as defined by
\begin{eqnarray}\label{eq:chi2}
    \chi^2 = \sum_i\frac{(O_{i}^\mathrm{obs}-O^\mathrm{mod}_{i})^2}{(\sigma^{\rm obs}_{i}+\sigma^{\rm mod}_{i})^2}
\end{eqnarray}
where the minimization is performed using the Optim.jl package~\cite{Mogensen2018}. 
The data set $O_{i}^\mathrm{obs}$ in the fit includes experimental data of both mass spectra and decay constants. 
Additionally, we incorporate data from lattice QCD simulations for the decay constants~\cite{McNeile:2012qf, Davies:2010ip, Donald:2012ga, Colquhoun:2014ica}.
When minimizing $\chi^2$ to fit the data, it is also necessary to optimize the Gaussian parameter for each iteration in the $\chi^2$ fit to ensure the minimum of the energy. 
This is essential to satisfy the variational principle, as otherwise, the results are not valid due to not reaching the energy minimum.

\begin{figure}[b]
	\centering
	\includegraphics[width=0.9\columnwidth]{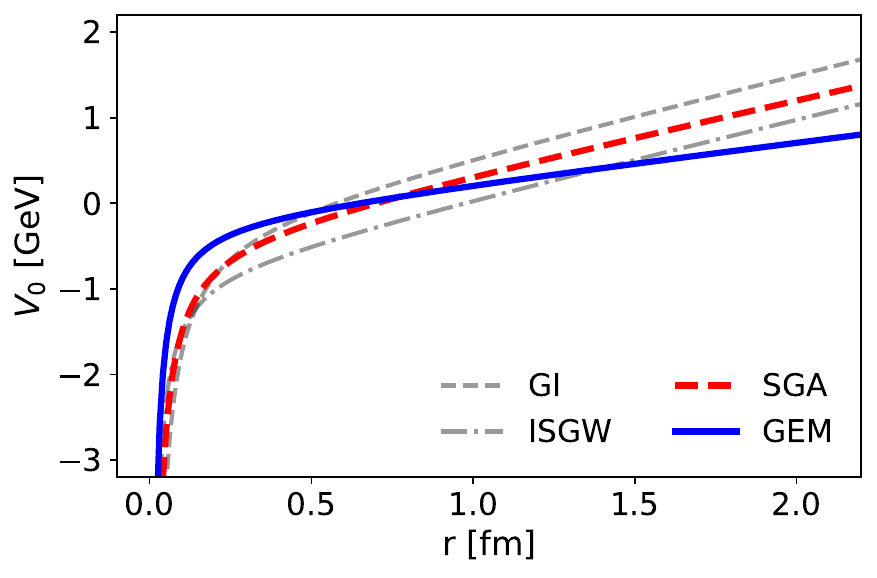}
 	\caption{\label{fig:potential} Spatial dependence of the confinement and color Coulomb potentials for the SGA and GEM using the model parameters in Table~\ref{tab:parameter}. 
    The potentials are comparable with those of the GI~\cite{Godfrey:1985xj} and ISGW potentials~\cite{ Isgur:1988gb}.} 
\end{figure}

Here, we adjust the model error $\sigma^{\rm mod}_i = \sigma_\mathrm{err} O_{i}^\mathrm{obs}$, where $\sigma_\mathrm{err}$ is the percentage error of the $\mathcal{O}_i^{\rm obs}$, to get $\chi^2/N_\mathrm{dof}\approx 1$. 
The $N_\mathrm{dof}$ represents the number of degrees of freedom obtained by subtracting the number of free parameters $(N_\mathrm{par}=8)$ from the number of data points $(N_\mathrm{data}=32)$.
We set $\sigma_\mathrm{err}^{c}=2\%$, $\sigma_\mathrm{err}^{b}=1\%$, and $\sigma_\mathrm{err}^{f}=5\%$, for the model error of the mass of charmed mesons $(q\bar{c},s\bar{c},c\bar{c})$, bottom mesons $(q\bar{b},s\bar{b},c\bar{b},b\bar{b})$, and decay constant for all of the heavy mesons, respectively.
The inclusion of this additional $\sigma_{\rm mod}$ in the $\chi^2$ ensures an unbiased fit, especially when the data precision varies significantly~\cite{Garcia-Tecocoatzi:2022zrf}. 
Note that $\sigma_{\rm mod}$ becomes dominant when $\sigma_{\rm expt}$ is negligibly small. 
Furthermore, we use a relative $\sigma_{\rm mod}$ because mass spectra and decay constants differ by up to an order of magnitude.

It should be noted that the fitted parameter sets for SGA and GEM
are significantly different. 
Indeed, Fig.~\ref{fig:wf} shows that considering a fixed parameter set for both methods, the decay constants may differ by up to 30\%. 
This shows again that the two approaches are conceptually different.
Consequently, we need to perform a fit to the data independently for both methods.
In GEM, using 10 basis functions yields an accuracy of about 4 digits.

\section{Results and Discussion}
\label{sec:result}

In this section, we first present the obtained model parameters for SGA and GEM
from fitting to static properties such mass spectra and decay constants. 
Subsequently, we continue to discuss the difference between the 
two methods for LFWFs and structural observables.

\subsection{Model Parameters}

The results for the fitted model parameters for both SGA and GEM are presented in Table~\ref{tab:parameter}. 
Although the form of the Hamiltonian is the same for both methods, the model parameters obtained from the fit show differences.
The obtained mass spectra and decay constants for ground-state heavy mesons, 
both included in our fitting process, are shown in Fig.~\ref{fig:mass}. 
Although $\chi^2/N_\mathrm{dof}$ for the SGA is smaller, it is less then one for both methods with $\sigma_\mathrm{mod}$ ranging from 1-5\% as explained previously.
This shows that the SGA is simple yet powerful while the GEM can provide a more accurate WF but requires more effort to obtain a more refined model Hamiltonian to improve agreement with the data.

Initially, we attempted to fit with unconstrained parameters for SGA and GEM. In this setup, we found reasonable agreement with the data, but the obtained $m_q$ was rather small, around 60 MeV, which is smaller than the typical constituent quark mass $m_q$ = 200-350 MeV. 
Because of that, we tested other observables such as the EM radius with this parameter set, and the obtained radius for $D^+$ was around 0.480 fm$^2$, much larger than the lattice QCD data of around 0.150 fm$^2$ discussed in Sec.~\ref{sec:formfactor}. Furthermore, $m_q$ was more tightly constrained with the observables for light mesons, which were not included in the present work. Due to these observations, we set the lower bound for $m_q=220$ MeV to the commonly used value in Refs.~\cite{Godfrey:1985xj,Choi:2007yu} for our analysis.

\begin{figure}[b]
	\centering
	\includegraphics[width=0.95\columnwidth]{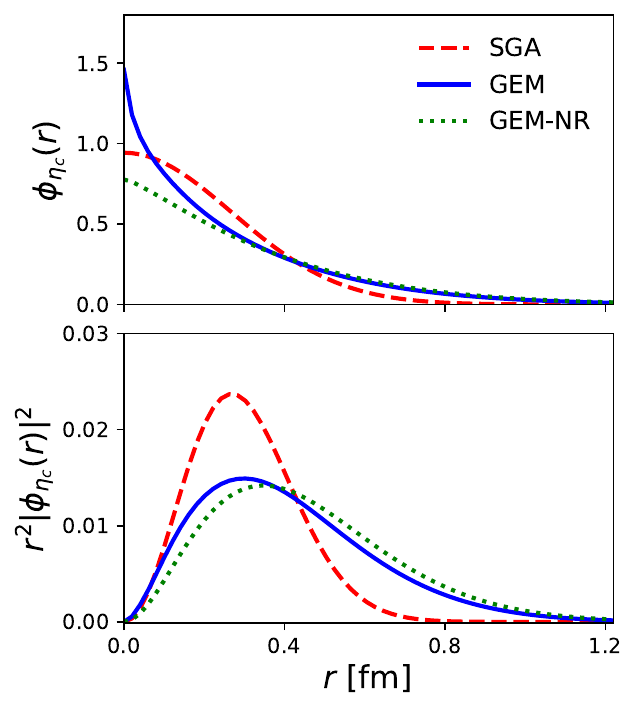}
 	\caption{\label{fig:wfcom} WF (upper panel) and the density (lower panel) of the WFs of $\eta_c(1S)$ for both SGA, GEM, and GEM-NR. The GEM and GEM-NR employ a relativistic and nonrelativistic kinetic energy, respectively. As compared to the one for the SGA, the WF for the GEM is more enhanced at short and long distances, but it is more suppressed at an intermediate distance. The WF near the origin for the GEM-NR is more suppressed than the one for the GEM.} 
\end{figure}

The spatial dependence of the potentials for the SGA and GEM are shown in Fig.~\ref{fig:potential} and compared with the Godfrey-Isgur (GI) model~\cite{Godfrey:1985xj} and Isgur-Scora-Grinstein-Wise (ISGW) with $\alpha_s$ = 0.3~\cite{Isgur:1988gb}.
While both potentials in this work are comparable with those of the literature,
one can see that the potential for the GEM is more enhanced at a short distance and suppressed at a long distance as compared to those for the SGA. 
It is worth noting that the outcome for GEM is notably influenced by variations in the model Hamiltonian, leading to distinct shape, as illustrated in Fig.~\ref{fig:potential}.
In this case, the enhancement of the WF at the origin for the GEM as shown in Fig.~\ref{fig:wf}, which is plausibly due to the use of relativistic kinetic energy, leads to the best fit with different potential parameters.

\subsection{Wave functions and their asymptotic behaviors}

In Fig. \ref{fig:wfcom} we show as an example the WF (upper panel) and density (lower panel) of $\eta_c$ for both SGA and GEM.
Unlike those obtained with the same parameters as shown in Fig.~\ref{fig:wf}, the WFs of both methods are now much closer to each other.
Nevertheless, we can identify some distinct differences, in particular, the WF for the GEM is more enhanced at the origin and extends to larger distances. 
If we replace the kinetic energy with the nonrelativistic one while keeping the parameters the same as those for the GEM, we obtain that the WF denoted as the GEM-NR is rather suppressed at the origin as shown in Fig.~\ref{fig:wfcom}.
This is mainly because the relativistic kinetic energy provides weaker repulsion near the origin than the nonrelativistic one.

\begin{figure}[t]
	\centering
	\includegraphics[width=0.9\columnwidth]{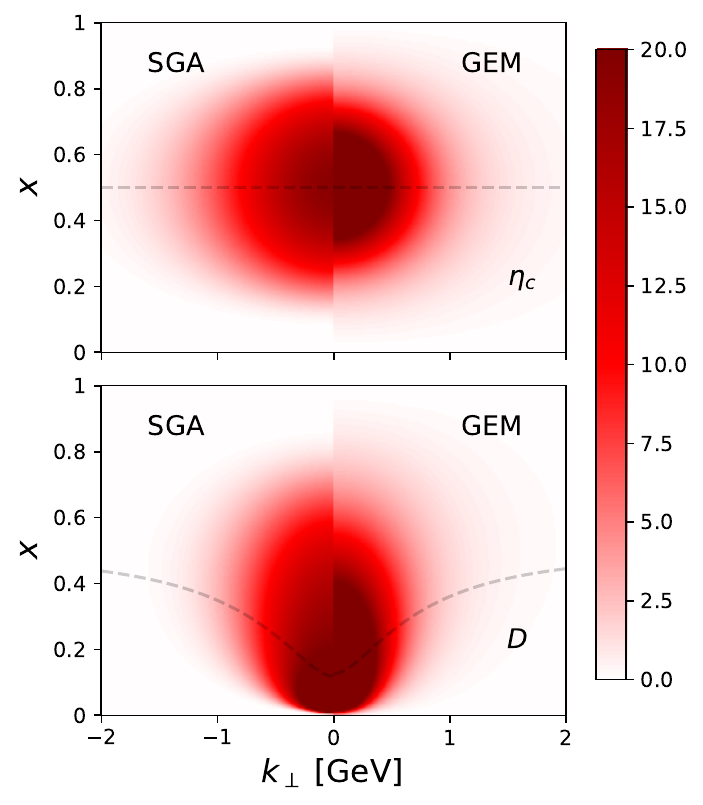}
 	\caption{\label{fig:lfwf} LFWF $(\varPsi_{\uparrow\downarrow}^{00}-\varPsi_{\downarrow\uparrow}^{00})/2$ of the $\eta_c$ (upper panel) and of the $D$ meson (lower panel) for both SGA (left half) and GEM (right half). The LFWFs for the GEM are more extended to both endpoints of $x$ but suppressed at the intermediate $x$ as compared to those for the SGA. The dashed line represents $k_z=0$ obtained from Eq.~\eqref{eq:k_z}. } 
\end{figure}

Moreover, we plot the LFWFs $(\varPsi_{\uparrow\downarrow}^{00}-\varPsi_{\downarrow\uparrow}^{00})/2$ of the $\eta_c$ and $D$ meson in the upper and lower panels of Fig.~\ref{fig:lfwf}, respectively, exemplifying the cases of equal and unequal mass of constituents. 
The peak for the $\eta_c$ LFWF, which corresponds to $(k_z,\bm{k}_\perp)=0$, is at $x=1/2$ but the peak for the $D$ meson LFWF is at $x=m_q/(m_q+m_{\bar{q}})$ with $x$ carried by the light quark. 
Note that the endpoints $x=0$ and $x=1$ correspond to $k_z=-\infty$ and $k_z=\infty$, respectively.
Evidently, the LFWF for the GEM extends more to both endpoints of $x$ and the larger $\bm{k}_\perp$ region, which we discuss in more detail in Sec. III. D.

Since the GEM accurately approximates the eigenstate of the Hamiltonian, while the SGA only captures the size of the WF, it is important to check the asymptotic behavior of the WF in two different limits: long distance ($r \to \infty$) and short distance ($r \to 0$). Of special interest is the short-distance behavior, which corresponds to the high-momentum behavior ($k \to \infty$), which is dictated by perturbative QCD.

\begin{figure}[b]
	\centering
	\includegraphics[width=0.9\columnwidth]{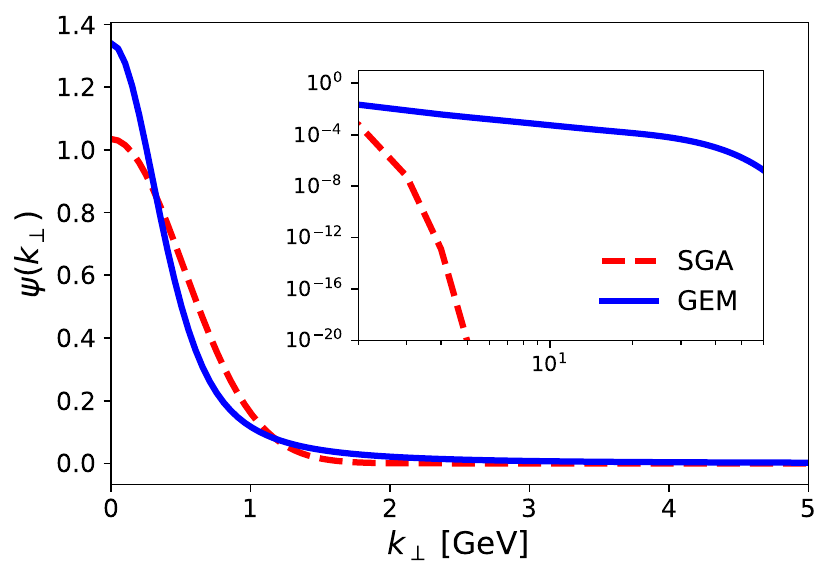}
 	\includegraphics[width=0.9\columnwidth]{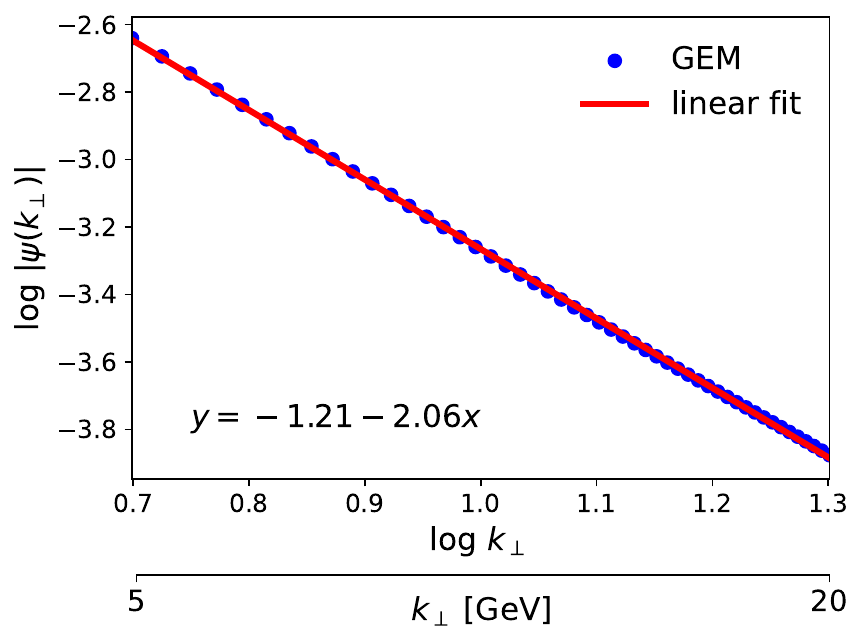}
 	\caption{\label{fig:asym_kt} (Upper panel) The $k_\perp$ dependence of the WF $\psi(k_\perp)$ where the $k_z$ component has been integrated. The inset shows a $k_\perp$ dependence up to 100 GeV in a logarithmic scale. (Lower panel) a linear fit to a WF $\psi(k_\perp)$ on a log-log scale. It implies that $\psi(k\to \infty)\propto 1/k_\perp^2.$} 
\end{figure}


In the long-distance limit $r\to \infty$, the Schrödinger equation for the $S$-wave with the power-law confinement reduces to
\begin{eqnarray}
 \frac{\nabla^2}{2\mu}\psi(r) = br^p\psi(r)
\end{eqnarray}
with the reduced quark mass $\mu$, the confinement parameter $b$, and exponent $p$. 
If we assume a decaying WF $\psi(r) \propto \exp(-Br^n)$, 
we can obtain the asymptotic parameters $n=(p+2)/2$ and $B=\sqrt{2\mu b/n^2}$.
While a HO confinement ($br^2$) results in a Gaussian asymptote,
the linear confinement ($br$) considered here leads to
\begin{equation}
    \psi_\text{Lin}(r\to\infty) \propto \exp(-\sqrt{\frac{8\mu b}{9}}\ r^{3/2}),
\end{equation}
which decays more slowly compared to the HO case, as expected.
Note that the Airy function is the eigenfunction of the linear potential~\cite{abramowitz-stegun}, 
and we reproduced here its asymptotic limit.


In the short distance, the dominant interaction is the color Coulomb potential, stemming from the one-gluon exchange and the relativistic effect becomes important as the relative momentum becomes large. As $r\to 0$, the Schrödinger equation is reduced to 
\begin{eqnarray}
 2\sqrt{(m^2-\nabla^2)}\psi(r) = \frac{4\alpha_s}{3r}\psi(r) 
\end{eqnarray}
where we consider the equal-mass case $m=m_q=m_{\bar{q}}$ for simplicity.
Near the origin, the WF has a power-law behavior $\psi(r\to 0)\to r^n$, which is also observed when using the Dirac Hamiltonian~\cite{Ito:1987nt}. 
Such behavior is well approximated by the GEM, as shown in Fig.~\ref{fig:wfcom}, 
but differs significantly in the SGA. 
Additionally, this asymptotic WF is divergent, but the truncation of the number of basis functions in the GEM regulates it.

\begin{table}[b]
	\begin{ruledtabular}
		\caption{Numerical results of mass spectra [MeV] of the ground state of heavy mesons for both SGA and GEM compared with the experimental data in Ref.~\cite{ParticleDataGroup:2022pth}.  } 
		\label{tab:mass}
		\begin{tabular}{cccc}
              & SGA & GEM & Expt. \\ \hline
        $M_D$       & 1909 & 1916 & 1869.66(05)   \\
        $M_{D^*}$   & 1992 & 1990 & 2010.26(05)  \\
        $M_{D_s}$   & 1988 & 2001 & 1968.35(7)   \\
        $M_{D_s^*}$ & 2064 & 2065 & 2112.2(4)    \\
        $M_{\eta_c}$ & 3012 & 3019 & 2983.9(4)    \\
        $M_{J/\psi}$ & 3066 & 3059 & 3096.900(6)   \\ 
        $M_B$       & 5290 & 5268 & 5279.34(12)     \\
        $M_{B^*}$   & 5325 & 5298 & 5324.70(21)     \\
        $M_{B_s}$   & 5356 & 5342 & 5366.88(14)   \\
        $M_{B_s^*}$ & 5390 & 5369 & $5415.4^{+1.8}_{-1.5}$   \\
        $M_{B_c}$   & 6289 & 6297 & 6274.47(32)   \\
        $M_{B_c^*}$ & 6325 & 6323 & \dots         \\
        $M_{\eta_b}$ & 9420 & 9488 & 9398.7(2.0)    \\
        $M_{\Upsilon}$ & 9459 & 9515 & 9460.30(26) \\ 
		\end{tabular}
	\end{ruledtabular}
\end{table}

To explicitly demonstrate a power-law behavior in the model calculation, 
it is more effective to examine the WF in momentum space, as this reveals the behavior more clearly. 
The momentum-space WF is obtained via Fourier transformation as given by 
\begin{eqnarray}
    \psi(k) &=& \int  \psi(r)\ \mathrm{e}^{-i \bm{k}\cdot \bm{r}} \dd^3r 
    =  \frac{1}{ik} \int_0^\infty \mathrm{e}^{ikr}\ r^{n+1} \dd r, \nonumber\\
    &=&  \frac{1}{ik^{(n+3)}}\Gamma(n+2),
\end{eqnarray}
where we have used the plane-wave expansion for the $S$ wave. 
We obtain 
\begin{eqnarray}
    \psi(k\to \infty) \propto \frac{1}{k^{(n+3)}}.
\end{eqnarray}
Figure~\ref{fig:asym_kt} shows the WF integrated over $k_z$, 
\begin{eqnarray}
    \psi(k_\perp) = \int_0^\infty \dd k_z\ \psi(k_z,k_\perp),
\end{eqnarray}
for $\eta_c$ and as $k\to\infty$.
From the inset of the upper panel, it is evident that the GEM gives a linear dependence up to about $k_\perp=50$ GeV, 
which is eventually broken at higher momenta.
This is because the GEM uses Gaussian basis functions, and the accuracy is limited by the narrowest (widest) basis function in position (momentum) space, and the number of basis functions. 
Our fit of the asymptotic WF between $k_\perp=$ 5-20 GeV on a log-log scale reveals a linear relationship with a slope of approximately $-2$. 
This corresponds to a damping factor of $1/k_\perp^2$, which also implies that the $\psi(r\to 0)\propto 1/r.$
This finding is consistent with the predictions of perturbative QCD~\cite{Ji:2003yj} and the Bethe-Salpeter method~\cite{Shi:2018zqd}.

\subsection{Mass Spectra}

The mass spectra for the ground state of heavy mesons obtained by the SGA and GEM are presented in Table~\ref{tab:mass}. 
Both methods yield similar results which are consistent with the experimental data of the Particle Data Group~\cite{ParticleDataGroup:2022pth}.
Since the mass spectra originate from the average of short- and long-distance effects of the WFs, a similar result from both methods is expected.

Here we use different $\sigma_i^{\mathrm{mod}}$ for mesons containing the charm ($\sigma_\mathrm{err}^c=2\%$) and bottom quark ($\sigma_\mathrm{err}^b=1\%$) to obtain a better fit with the data for all heavy mesons. 
If we employ the same $\sigma_\mathrm{err}=2\%$, the results for heavier mesons, such as bottomonia, will be less accurate. 
On the other hand, if we use absolute values such as $\sigma_{\rm mod}=50$ MeV, the results for the heavy-light meson become less accurate. 
The use for the relative error in $\sigma^{\rm mod}$ becomes important if we include light mesons in the fit since their masses are much smaller.
Furthermore, in this work, we analyze not only the mass spectra but also other observables. 
In this case, the results should have reasonable agreement with the data for other observables as well. 
If we fit only the mass spectra, the prediction for decay constants can be bad.
Therefore, the mass spectra provide a necessary, but not sufficient condition to constrain the WF.

\subsection{Decay Constants}

The decay constants obtained for both SGA and GEM are tabulated in Table~\ref{tab:cons}, and compared with experimental data, extracted from the leptonic and weak decays~\cite{ParticleDataGroup:2022pth}, and lattice QCD data~\cite{McNeile:2012qf, Davies:2010ip, Donald:2012ga, Colquhoun:2014ica}. 
Note that there are some discrepancies between both lattice QCD and experimental data, because of which we employ $\sigma_\mathrm{err}^f=5\%$ in the fit.
Since there is no experimental data for $f_V^\perp$, we only compare our results with other theoretical model~\cite{Hwang:2010hw}.

We find that the results for both methods have reasonable agreement with the data as shown in Table~\ref{tab:cons}.
Although in the nonrelativistic limit the decay constant is related to the WF at the origin via the famous Van Royen-Weisskopf formula~\cite{van1967hardon}, 
the results for both methods are comparable once the range parameters of the WFs are fitted to the data.
Therefore, it is of great interest to analyze momentum-dependent quantities, instead of constant observables, to unveil the difference between the two methods.

While $f_{\rm P} < f_{\rm V}^\parallel$ is always held if the spin-spin interaction is treated perturbatively~\cite{Arifi:2022pal}, $f_{\rm P}$ can be larger than $f_{\rm V}^\parallel$ when the spin-spin interaction is treated nonperturbatively and the smearing parameter $\tilde{\Lambda}$ plays a crucial role in determining the hierarchy as discussed previously~\cite{Choi:2015ywa}. For instance, $f_{\eta_b}=691$ MeV $> f_\Upsilon^\parallel=688$ MeV is obtained in SGA as shown in Table~\ref{tab:mass}.
In GEM, $f_{\eta_b}<f_{\Upsilon}^\parallel$ is observed despite the quark mass dependence added to the $\tilde{\Lambda}$ parameter. 
At the moment, lattice QCD and experimental data for $f_\Upsilon^\parallel$ have some discrepancy.
More precise data is therefore desirable to resolve the hierarchy which is useful to further constrain the WF.

Furthermore, from Eq.~\eqref{eq:cons}, we see that $\mathcal{O}_{\rm P} = 2\mathcal{O}_{\rm V}^\perp - \mathcal{O}_{\rm V}^\parallel$ resulting in $f_{\rm P} = 2f_{\rm V}^\perp - f_{\rm V}^\parallel$ if the radial WFs of pseudoscalar and vector mesons are the same~\cite{Arifi:2022pal}. 
Here we find that $f_{\rm P} > 2f_{\rm V}^\perp - f_{\rm V}^\parallel$ when the spin-spin interaction is treated nonperturbatively and it applies to both SGA and GEM.

\begin{table}[t]
	\begin{ruledtabular}
		\caption{Numerical results of decay constants [MeV] of ground-state heavy mesons for both SGA and GEM, and compared with experimental~\cite{ParticleDataGroup:2022pth} as well as lattice QCD data~\cite{McNeile:2012qf, Davies:2010ip, Donald:2012ga, Colquhoun:2014ica}. We also compare the results for $f_{\rm V}^\perp$ by Ref.~\cite{Hwang:2010hw}. } 
		\label{tab:cons}
		\begin{tabular}{cccccc}
                                & SGA & GEM & Lattice QCD & Expt. & \cite{Hwang:2010hw} \\ \hline
        $f_D$                   & 224 & 225 & 211(14) & 206.7(8.9) & \dots  \\
        $f_{D^*}^\parallel$     & 251 & 249 & 245(20) & \dots  & \dots   \\
        $f_{D^*}^\perp$         & 227 & 213 &  \dots  &  \dots & 233 \\
        $f_{D_s}$               & 253 & 249 & 231(12) & 257.5(6.1) & \dots   \\
        $f_{D_s^*}^\parallel$   & 276 & 268 & 272(16) & \dots & \dots  \\
        $f_{D_s^*}^\perp$       & 252 & 233 &  \dots &  \dots & 303  \\
        $f_{\eta_c}$            & 376 & 355 & 394.7(2.4) & 335(75) & \dots    \\
        $f_{J/\psi}^\parallel$  & 384 & 378 & 405(6) & 407(5) & \dots   \\ 
        $f_{J/\psi}^\perp$      & 363 & 337 & \dots &  \dots  & \dots  \\
        $f_B$                   & 200 & 213 & 179(18) & 188(25) & \dots     \\
        $f_{B^*}^\parallel$     & 207 & 240 & 196(24) & \dots  & \dots  \\
        $f_{B^*}^\perp$         & 198 & 213 & \dots &  \dots  & 214  \\
        $f_{B_s}$               & 229 & 238 & 204(16) & \dots & \dots    \\
        $f_{B_s^*}^\parallel$   & 233 & 263 & 229(20) & \dots & \dots  \\
        $f_{B_s^*}^\perp$       & 224 & 235 & \dots   &  \dots & 297   \\
        $f_{B_c}$               & 426 & 412 & $427^{+6}_{-2}$ & \dots & \dots  \\
        $f_{B_c^*}^\parallel$   & 423 & 449 & \dots & \dots & \dots   \\
        $f_{B_c^*}^\perp$       & 409 & 405 & \dots & \dots &374 \\
        $f_{\eta_b}$             & 692 & 659 & $667^{+6}_{-2}$ & \dots & \dots   \\
        $f_{\Upsilon}^\parallel$ & 688 & 729 & 649(31) & 715(5) & \dots  \\
        $f_{\Upsilon}^\perp$     & 668 & 660 & \dots & \dots & \dots   \\
		\end{tabular}
	\end{ruledtabular}
\end{table}

\begin{figure*}[t]
	\centering
     \includegraphics[width=1.8\columnwidth]{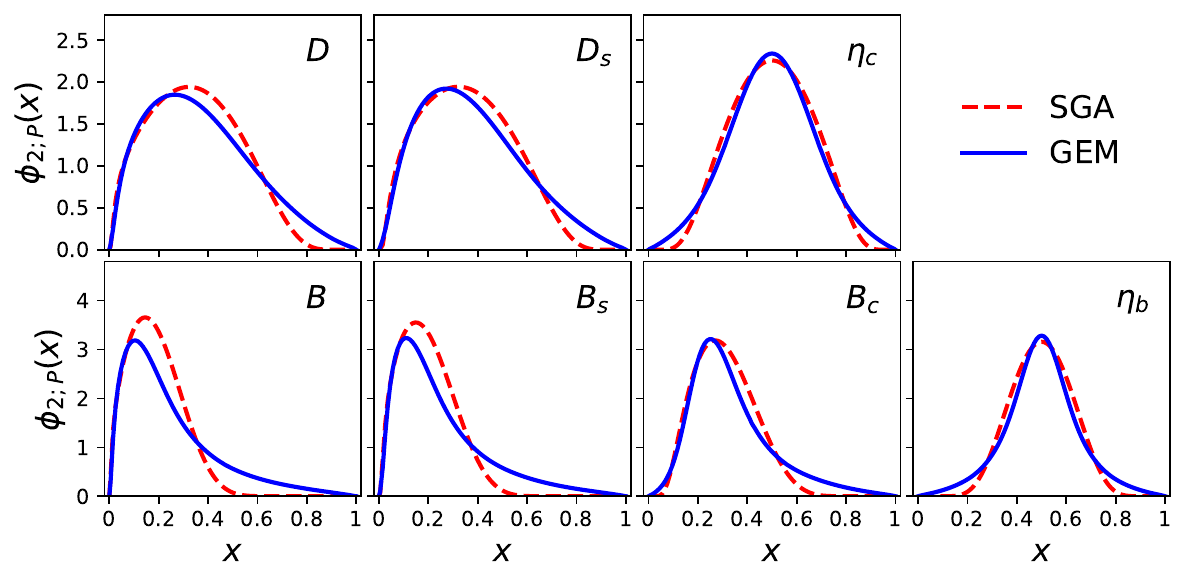}
 	\caption{\label{fig:daheavy} Twist-2 DAs $\phi_{2;P}(x)$ of pseudoscalar heavy mesons with various quark flavor contents for both SGA (red) and GEM (blue). Compared to the results for the SGA, we find the DAs for the GEM are more enhanced near both endpoints. } 
\end{figure*}

\begin{table*}[t]
	\begin{ruledtabular}
		\caption{Six lowest $\xi$-moments of $\phi_{2;P}(x)$ for both SGA and GEM, where $\xi=2x-1$. The $\xi$ moments for the GEM are generally larger than those for the SGA, except for the even $\xi$ moments of $B$ and $B_s$ mesons.}
		\label{tab:xi-ps}
		\begin{tabular}{crrrrrrrrrrrrrr} 
             & \multicolumn{2}{c}{$\phi_{2;D}$} & \multicolumn{2}{c}{$\phi_{2;D_s}$} & \multicolumn{2}{c}{$\phi_{2;\eta_c}$} & \multicolumn{2}{c}{$\phi_{2;B}$} & \multicolumn{2}{c}{$\phi_{2;B_s}$} & \multicolumn{2}{c}{$\phi_{2;B_c}$} & \multicolumn{2}{c}{$\phi_{2;\eta_b}$} \\ 
                             & SGA      & GEM     & SGA     & GEM     & SGA    & GEM     & SGA     & GEM      & SGA     & GEM       & SGA     & GEM      & SGA     & GEM \\ \hline
            $\expval{\xi^1}$ & $-0.303$ & $-0.262$& $-0.280$& $-0.244$& $\dots$& $\dots$ & $-0.629$& $-0.508$ & $-0.609$& $-0.497$  & $-0.381$& $-0.329$ & $\dots$ & $\dots$\\ 
            $\expval{\xi^2}$ & $0.218$  & $0.238$ & $0.203$ & $0.222$ & $0.099$& $0.118$ & $0.438$ & $0.399$  & $0.415$ & $0.385$   & $0.202$ & $0.216$  & $0.056$ & $0.086$\\ 
            $\expval{\xi^3}$ & $-0.133$ & $-0.117$& $-0.117$& $-0.100$& $\dots$& $\dots$ & $-0.324$& $-0.279$ & $-0.303$& $-0.263$  & $-0.115$& $-0.104$ & $\dots$ & $\dots$\\ 
            $\expval{\xi^4}$ & $0.103$  & $0.112$ & $0.089$ & $0.098$ & $0.023$& $0.039$ & $0.251$ & $0.234$  & $0.231$ & $0.218$   & $0.071$ & $0.080$  & $0.008$ & $0.026$\\ 
            $\expval{\xi^5}$ & $-0.076$ & $-0.067$& $-0.063$& $-0.053$& $\dots$& $\dots$ & $-0.201$& $-0.180$ & $-0.182$& $-0.164$  & $-0.046$& $-0.043$ & $\dots$ & $\dots$\\ 
            $\expval{\xi^6}$ & $0.062$  & $0.066$ & $0.050$ & $0.055$ & $0.007$& $0.018$ & $0.164$ & $0.157$  & $0.147$ & $0.141$   & $0.031$ & $0.038$  & $0.002$ & $0.012$\\ 
		\end{tabular}
	\end{ruledtabular}
\end{table*}

\subsection{Twist-2 Distribution Amplitudes}

In Fig.~\ref{fig:daheavy}, we show the leading-twist DAs of pseudoscalar mesons for both SGA and GEM. 
Overall, the results are comparable, but they exhibit different behaviors near both endpoints. 
For heavy-light mesons, the difference is more evident near $x=1$. 
For instance in the case of the $B$ meson, the DAs in the two methods look quite different, and these features are commonly observed in other mesons. 

The $\phi_{2;B}(x)$ for the GEM is more suppressed in the region of $0.1<x<0.3$ and more enhanced in the region of $x>0.3$. 
While the $\phi_{2;B}(x)$ in GEM is more extended to the $x>0.5$ region, the $\phi_{2;B}(x)$ for the SGA is concentrated in the $x<0.5$ region. 
The endpoint behavior at $x=1$ is related to the high-momentum part of the WF and accordingly a short-distance part of the WF. 
For the SGA, this suppression can be directly inferred from the WF which is rather suppressed at a short distance as shown in Fig.~\ref{fig:wfcom}. 
On the other hand, the WFs for the GEM are enhanced at a short distance. 

\begin{figure}[t]
	\centering
 	\includegraphics[width=0.87\columnwidth]{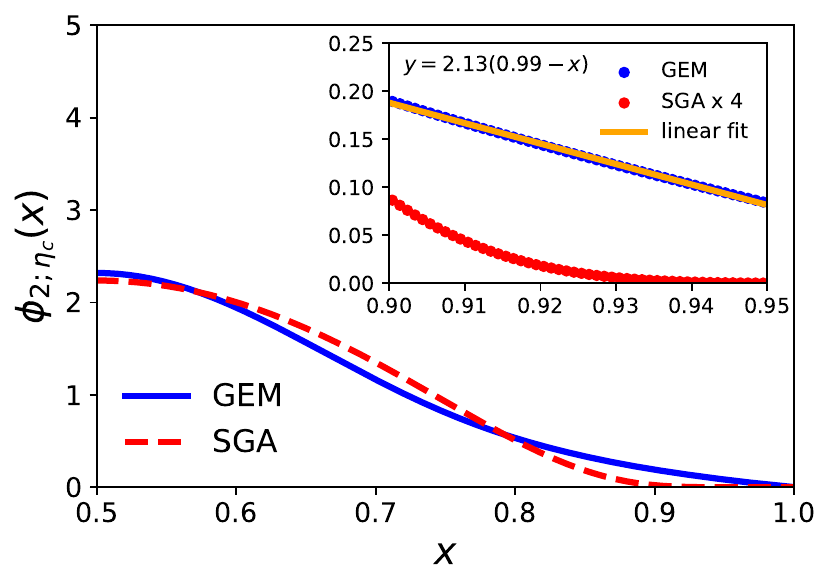}
 	\caption{\label{fig:asym_da} DAs and their endpoint behavior $\phi_{2;\eta_c}(x\to 1)$ in two methods. The inset displays the DA near $x=1$, with a behavior $\phi(x\to1)\propto (1-x)$ obtained for the GEM, while that for the SGA behaves differently. } 
\end{figure}

We also plot the asymptotic behavior of the $\phi_{2;\eta_c}(x)$ in Fig.~\ref{fig:asym_da}.
It is evident that the DA near $x=1$ for the GEM shows a linear dependence as
\begin{eqnarray}
\phi_{2;\eta_c}(x\to1) = 2.13(0.99-x),
\end{eqnarray}
while it behaves differently in the case of SGA.
The linear behavior is needed to yield a meson PDF of $(1-x)^2$ near $x=1$, as predicted by perturbative QCD~\cite{Farrar:1975yb}.
It is worth noting that we fit the DA between $0.9 \leq x \leq 0.95$. as they eventually start to deviate from the linear form when it is much closer to $x=1$, similar to Fig.~\ref{fig:asym_kt}. 
In the BLFQ approach~\cite{Li:2017mlw}, this asymptotic WF is used as a basis function such as $\varPsi_L(x)\propto x^\alpha(1-x)^\beta$. 
This produces the asymptotic behavior of DAs by construction, but it may also break the spherical symmetry.

\begin{figure*}[t]
	\centering
	\includegraphics[width=1.8\columnwidth]{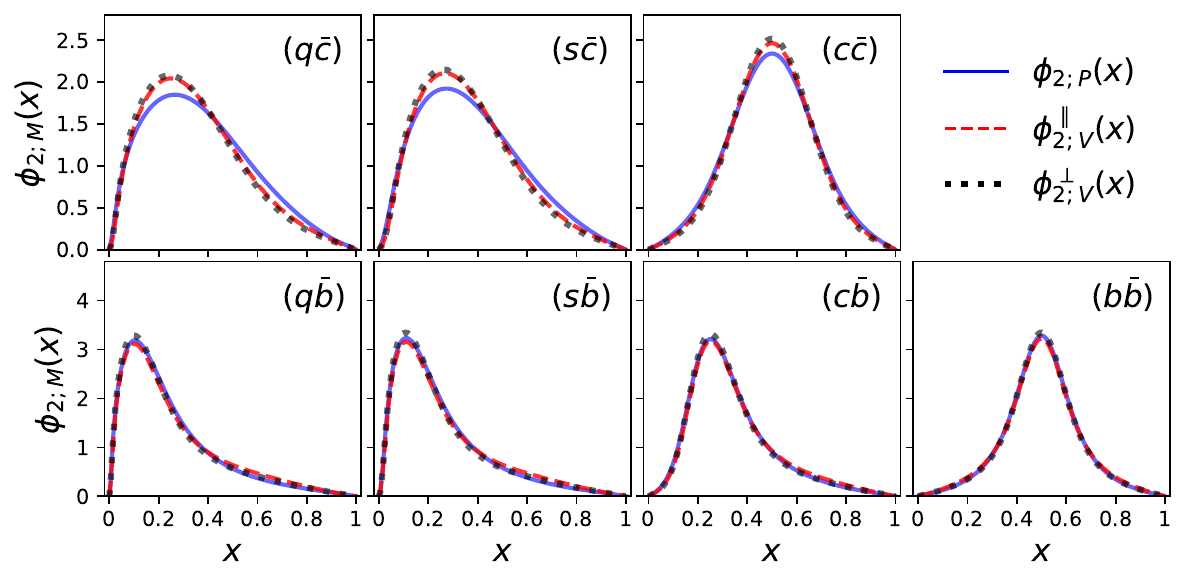}
 	\caption{\label{fig:dapvvt} Twist-2 DAs of pseudoscalar $\phi_{2;P}(x)$ and vector heavy mesons with longitudinal $\phi_{2;V}^\parallel(x)$ and transverse $\phi_{2;V}^\perp(x)$ polarization. Although we include the spin-spin term nonperturbatively, the differences in the DAs of pseudoscalar and vector mesons are rather small. Only the mesons with charm quark show some visible difference in the DAs. } 
\end{figure*}

\begin{table*}[t]
	\begin{ruledtabular}
		\caption{Six lowest $\xi$ moments of $\phi_{2;V}^\parallel(x)$ and $\phi_{2;V}^\perp(x)$ obtained with the GEM, where $\xi=2x-1$.}
		\label{tab:xi-vec}
		\begin{tabular}{crrrrrrrrrrrrrr} 
                    & $\phi_{2;D^*}^\parallel$ & $\phi_{2;D^*}^\perp$ 
                    & $\phi_{2;D_s^*}^\parallel$ & $\phi_{2;D_s^*}^\perp$ 
                    & $\phi_{2;J/\psi}^\parallel$ & $\phi_{2;J/\psi}^\perp$ 
                    & $\phi_{2;B^*}^\parallel$ & $\phi_{2;B^*}^\perp$ 
                    & $\phi_{2;B_s^*}^\parallel$ & $\phi_{2;B_s^*}^\perp$ 
                    & $\phi_{2;B_c^*}^\parallel$ & $\phi_{2;B_c^*}^\perp$ 
                    & $\phi_{2;\Upsilon}^\parallel$ & $\phi_{2;\Upsilon}^\perp$ \\ \hline 
            $\expval{\xi^1}$ & $-0.295$ & $-0.311$& $-0.272$& $-0.286$& $\dots$& $\dots$& $-0.473$& $-0.501$ &$-0.467$ & $-0.494$ & $-0.308$ & $-0.326$ & $\dots$ & $\dots$\\ 
            $\expval{\xi^2}$ & $0.242$  & $0.248$ & $0.219$ & $0.223$ & $0.109$& $0.105$& $0.386$ & $0.403$  &$0.374$  & $0.390$  & $0.213$  & $0.216$  & $0.092$ & $0.087$\\ 
            $\expval{\xi^3}$ & $-0.125$ & $-0.133$& $-0.105$& $-0.111$& $\dots$& $\dots$& $-0.263$& $-0.280$ &$-0.250$ & $-0.266$ & $-0.096$ & $-0.101$ & $\dots$ & $\dots$\\ 
            $\expval{\xi^4}$ & $0.112$  & $0.117$ & $0.094$ & $0.096$ & $0.034$& $0.032$& $0.225$ & $0.238$  &$0.210$  & $0.222$  & $0.078$  & $0.079$  & $0.029$ & $0.027$\\ 
            $\expval{\xi^5}$ & $-0.069$ & $-0.074$& $-0.053$& $-0.056$& $\dots$& $\dots$& $-0.170$& $-0.182$ &$-0.155$ & $-0.166$ & $-0.039$ & $-0.041$ & $\dots$ & $\dots$\\ 
            $\expval{\xi^6}$ & $0.065$  & $0.068$ & $0.051$ & $0.052$ & $0.016$& $0.015$& $0.150$ & $0.159$  &$0.135$  & $0.143$  & $0.036$  & $0.036$  & $0.014$ & $0.013$\\ 
		\end{tabular}
	\end{ruledtabular}
\end{table*}

Furthermore, a previous study~\cite{Arifi:2023jfe} has shown that the DAs using SGA could not reproduce the enhancement near the endpoints seen in lattice QCD data~\cite{LatticeParton:2022zqc}. 
This clearly shows the limitations of the SGA due to the fixed form of the WF. 
These observations suggest the use of the eigenfunction of Hamiltonian instead of a single Gaussian Ansatz. 
In contrast, the superposition of Gaussian basis functions with different range parameters in GEM can produce a general WF with richer features.
Furthermore, it is worth noting that the enhancement near the endpoints of DAs can alternatively be obtained using the power-law Ansatz~\cite{Hwang:2010hw}, which is an asymptotic WF in the high momentum region.
However, this power-law Ansatz has also some limitations in its shape and has other problems such as convergence issues.

In Table~\ref{tab:xi-ps}, we tabulate the six lowest $\xi$-moments for the pseudoscalar DAs computed within the SGA and GEM. 
We find that the $\xi$ moments for the GEM are generally a bit larger than those from the SGA.
Moreover, the $\xi$ moments for mesons containing a bottom quark have larger deviations between both models. 
Such a large difference is reflected in the DAs as shown in Fig.~\ref{fig:daheavy}.
We also compare the $\xi$ moments with the SGA and power-law ansatz computed in Ref.~\cite{Hwang:2010hw}. 
For example, our computed odd $\xi$ moments for the $B$ meson, $\langle \xi^1 \rangle = -0.629 [-0.508]$ with SGA [GEM], are comparable to those in Ref.~\cite{Hwang:2010hw}, $\langle \xi^1 \rangle = -0.617 [-0.531]$ with SGA [Power-law].
This shows that our GEM results for the $\xi$ moment are more in line with those for the power-law ansatz.
This can be understood as they have more enhancements near $x =1$, which make them less asymmetrical with respect to $x=1/2$ and yield smaller absolute values of odd $\xi$ moments.

\begin{figure*}[t]
	\centering
	\includegraphics[width=1.9\columnwidth]{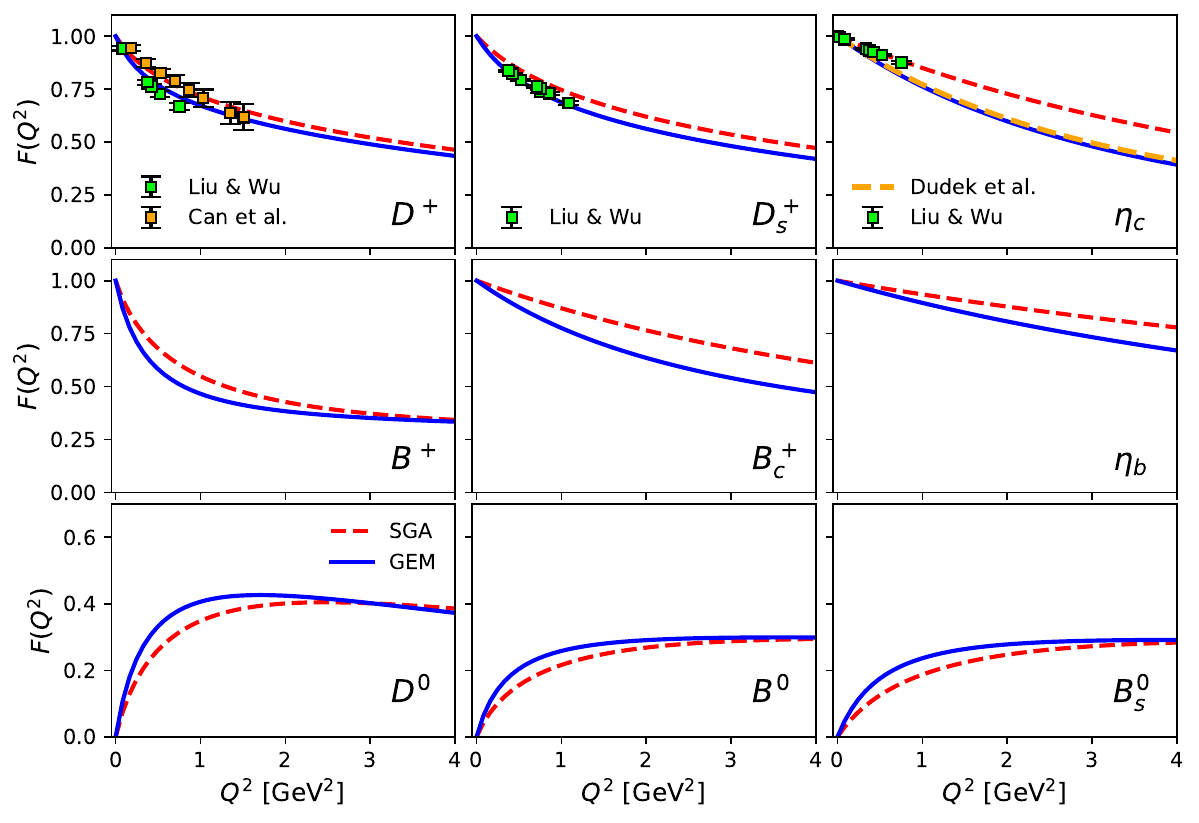}
 	\caption{\label{fig:formfactor} Computed EM form factor for pseudoscalar heavy mesons compared with available lattice QCD data~\cite{Can:2012tx, Li:2020gau, Li:2017eic, Dudek:2006ej}. Note that only the quark contribution is considered for $\eta_c$ and $\eta_b$, as otherwise the form factors vanish. } 
\end{figure*}

Since we now know the qualitative difference in the predictions of both methods, 
it is also interesting to compare the DAs for pseudoscalar and vector mesons obtained by the GEM. 
The comparisons are provided in Fig.~\ref{fig:dapvvt} where we plot the $\phi_{2;P}(x), \phi_{2;V}^\parallel(x),$ and $\phi_{2;V}^\perp(x)$.
The $\phi_{2;P}(x)$ are similar to the $\phi_{2;V}^{\parallel,\perp}(x)$ for bottom mesons, partly due to the heavy-quark symmetry, but there is some noticeable difference for the case of charm mesons. 
One can see that $\phi_{2;V}^{\parallel,\perp}(x)$ are generally comparable with each other, but have higher peaks as compared to $\phi_{2;P}(x)$.
When we use Gaussian basis functions with different range parameters (GEM), 
the resulting DAs always have a single peak. 
This is rather different than those obtained by expanding into the HO basis function where the mixture of higher $n$S can lead to oscillatory DAs~\cite{Dhiman:2019ddr} for the ground states such that the mixture is restricted to be very small~\cite{Arifi:2022pal}. 

For completeness, we also provide the six lowest $\xi$ moments for vector meson DAs in Table~\ref{tab:xi-vec}. 
The $\xi$ moments for the bottom vector and pseudoscalar mesons are rather similar, following their DAs as shown in Fig.~\ref{fig:dapvvt}.
For the charm mesons, the odd (even) $\xi$ moments for $\phi_{2,V}^{\parallel,\perp}$ are smaller (larger) than those of $\phi_{2,P}$ where the biggest difference is from the lowest odd $\expval{\xi^1}$ and even $\expval{\xi^2}$ moments.
While the even $\xi$ moments of $\phi_{2,V}^\perp$ are generally smaller than those of $\phi_{2,V}^\parallel$. For the charmonia and bottomonia,  the even (odd) $\xi$ moments of $\phi_{2,V}^\perp$ are larger (smaller) than those of $\phi_{2,V}^\parallel$ for the heavy-light mesons.

\begin{table}[b]
	\begin{ruledtabular}
	\caption{EM radius $\expval{r^2_\mathrm{EM}}$ of pseudoscalar mesons for both SGA and GEM, and compared with the lattice QCD~\cite{Can:2012tx, Li:2020gau, Li:2017eic, Dudek:2006ej} and the results of Refs.~\cite{Li:2017mlw,Hwang:2001th}. The results are given in units of fm$^2$. } 
		\label{tab:radius}
		\begin{tabular}{lccc}
                  & $D^+$  & ${D_s^+}$  & ${\eta_c}$  \\ \hline
        SGA       & 0.155 & 0.101 & 0.039  \\
        GEM       & 0.221 & 0.154 & 0.067 \\ 
        Lattice, Can et al.~\cite{Can:2012tx}  & 0.152(26) & \dots & \dots  \\
        Lattice, Li and Wu ~\cite{Li:2020gau, Li:2017eic}  & 0.176(69) & 0.125(13) & 0.052(4)  \\
        Lattice, Dudek et al.~\cite{Dudek:2006ej} & \dots & \dots & 0.063(1) \\ \hline  
                  & $B^+$  & ${B_c^+}$ & ${\eta_b}$  \\ \hline  
        SGA       & 0.297 &  0.034 & 0.008  \\
        GEM       & 0.459 &  0.067 & 0.017 \\ 
        BLFQ~\cite{Li:2017mlw}   & \dots & \dots &  0.012 \\ \hline
                  & $D^0$  & ${B^0}$ & ${B_s^0}$  \\ \hline  
        SGA       & $-0.252$ & $-0.161$ & $-0.099$  \\
        GEM       & $-0.377$ &  $-0.228$ & $-0.151$ \\ 
        LFQM~\cite{Hwang:2001th} & $-0.304$ & $-0.187$ & $-0.119$ \\
        LFQM, HQS limit~\cite{Hwang:2001th} & $-0.496$ & $-0.248$ & $-0.181$ \\
	   \end{tabular}
	\end{ruledtabular}
\end{table}

\subsection{Electromagnetic Form Factors}
\label{sec:formfactor}

The results of the EM form factors of the pseudoscalar heavy mesons for both SGA and GEM are presented in Fig.~\ref{fig:formfactor}, together with the lattice QCD~\cite{Can:2012tx, Li:2020gau, Li:2017eic, Dudek:2006ej}. 
We find that the results for $D^+$ and $D_s^+$ in both methods are comparable and consistent with the lattice QCD~\cite{Li:2017eic, Can:2012tx}. 
For the form factor of the $\eta_c$, both methods seem consistent with Ref.~\cite{Li:2020gau}, but only the result for the GEM can reproduce the lattice QCD (orange dashed line) of Dudek et al.~\cite{Dudek:2006ej}.
Note that we consider only the quark contribution for the case of $\eta_c$ and $\eta_b$, otherwise, the form factor vanishes. 

We see that the fall-off of EM form factors in low-$Q^2$ region for the GEM are generally faster than those for the SGA.
Even so, the difference between them may depend on the fit as the form factor depends on the quark masses and the potential parameters, where typically the smaller quark masses produce a faster fall-off of the form factor.

The mean squared of the charge radii from the form factors are shown in Table~\ref{tab:radius}, indicating that the obtained radii for the GEM are generally larger compared to those for the SGA.
This can be understood from the density of the WF plotted in Fig.~\ref{fig:wfcom}. 
Since the WF for the GEM extends more to long distances, the expected radius of $\expval{r^2_{\rm EM}}$ is larger than those for the SGA. 
Nevertheless, the obtained radii in both models are consistent with current lattice QCD~\cite{Can:2012tx, Li:2020gau, Li:2017eic, Dudek:2006ej}.
In particular, for the $B^+$ meson, the results from the two methods show a large deviation, i.e., $\expval{r_\mathrm{EM}^2}=0.297(0.459)$ fm$^2$ for the SGA (GEM), respectively. 
Therefore, more lattice QCD data on this observable is necessary to further constrain the models.

\begin{figure}[t]
	\centering
 	\includegraphics[width=0.95\columnwidth]{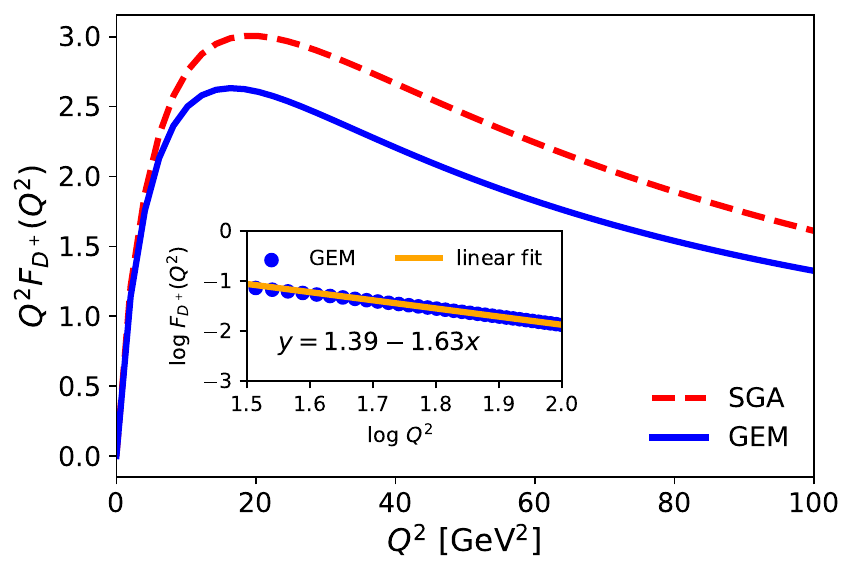}
 	\caption{\label{fig:asym_ff} EM form factor of the $D^+$ meson in the wide range of transfer momentum $Q^2$. We perform a linear fit to the form factor $F(Q^2)$ between $Q^2=$ 60-100 GeV$^2$ on a log-log scale. } 
\end{figure}

It is also crucial to examine the fall-off of the form factor in the high-$Q^2$ region, as it is dictated by perturbative QCD. 
As a demonstration, in Fig.~\ref{fig:asym_ff}, 
our calculations show that $Q^2F_{D^+}(Q^2)$ decreases with increasing $Q^2$ rather than remaining constant. 
Although there are logarithmic corrections $\mathrm{ln}\ Q^2$ due to the running of the strong coupling constant $\alpha_s(Q^2)$~\cite{Lepage:1979zb} that affect the $Q^2$ dependence, 
there could be other contributing factors. 
Then, we fitted the form factor $Q^2F_{D^+}(Q^2)$ using GEM for $Q^2$ values between 60 and 100 GeV$^2$ on a log-log scale as illustrated in Fig.~\ref{fig:asym_ff}. 
The results show that $F(Q^2) \propto 1/(Q^2)^{n}$ with $n = 1.63$, as indicated in the inset. 
When the form factor is adjusted by multiplying (dividing) by $\ln Q^2$, the exponent becomes $n = 1.40$ ($n = 1.86$), deviating from the perturbative QCD prediction of $n = 1$. 
In a previous study of the pion form factor~\cite{Maris:1998hc}, it was suggested that including the pseudovector component in the meson vertex using the Bethe-Salpeter approach, which is not considered in this work, could yield the correct asymptotic behavior. 
Furthermore, some nonperturbative effects may also give different $Q^2$ dependence in the form factor at high-$Q^2$ region~\cite{JointPhysicsAnalysisCenter:2024znj}. 
Further investigation into this matter is essential to fully understand the underlying mechanisms.

\section{Conclusion and Outlook} 
\label{sec:conclusion}

We have investigated the structure of heavy mesons using the single Gaussian Ansatz (SGA) and Gaussian expansion method (GEM) within the Light-Front Quark Model (LFQM).
To accomplish this, we have concentrated our efforts on the ground state of heavy mesons and investigate not only their static properties, but also their structural properties.
To determine the model parameters, we have performed a simultaneous $\chi^2$ fit to the static properties such as mass spectra and decay constants and examined the difference in the predictions of both methods especially in the structural properties such as DAs and EM form factor. 

We found that both methods yield similar static properties such as mass spectra and decay constants, and given the model uncertainty, they exhibit reasonable agreement with experimental and lattice QCD data.  
However, they show differences in the LFWFs and structural properties. 
In particular, the asymptotic behaviors of the WFs $\psi_0(k\to\infty)\propto 1/k_\perp^2$ and DAs $\phi(x\to 1)\propto (1-x)$ are correctly reproduced by the GEM, while they are not in the case of the SGA. 
These behaviors in the high-momentum region are governed by relativistic kinematics and Coulombic one-gluon exchange, which produce a power-law fall-off of the WF $\psi(r\to 0)\propto 1/r$.
Furthermore, the fall-off of the EM form factor in the low-$Q^2$ is faster for the GEM, giving better agreement when compared to the lattice QCD~\cite{Dudek:2006ej}, especially for the $\eta_c$ meson.

For future work, several directions can be explored:
first, it is crucial to investigate the form of the model Hamiltonian since the WF obtained in the GEM model is sensitive to the Hamiltonian, unlike in the SGA.  
Furthermore, expanding our calculations to include excited states and light mesons is of great importance in testing the applicability of GEM. 
Finally, our model can be tested on other form factors and hadron distributions.

\section*{Acknowledgement}

We would like to thank Atsushi Hosaka, Qi-Fang L\"{u}, Emiko Hiyama, Chueng-Ryong Ji, and Ho-Moeyng Choi for useful discussions.
The authors also thank the organizers of the Tohoku-RIKEN joint workshop 2023 where the present work was initiated.
We also thank the referees for bringing the important issues of asymptotic behaviors to our attention.
A.J.A. and L.H. were supported by the RIKEN Special Postdoctoral Researcher Program.
S. O. was supported by the RIKEN Junior Research Associate Program.
M. O. acknowledges the support by JSPS KAKENHI Grant Numbers, JP19H05159, and JP23K03427.

\bibliography{reference}

\end{document}